\documentclass[aps]{revtex4}
\usepackage{amssymb,epsf}

\begin{document}

\title{Phase transition and thermodynamic geometry of\\
Einstein-Maxwell-dilaton black holes}
\author{S. H. Hendi$^{1,2}$\footnote{
email address: hendi@shirazu.ac.ir}, A. Sheykhi$^{1,2}$\footnote{
email address: asheykhi@shirazu.ac.ir}, S. Panahiyan$^{1}$\footnote{%
email address: ziexify@gmail.com} and B. Eslam Panah$^{1}$\footnote{%
email address: behzad$_{-}$eslampanah@yahoo.com}}
\affiliation{$^1$ Physics Department and Biruni Observatory,
College of Sciences, Shiraz
University, Shiraz 71454, Iran\\
$^2$ Research Institute for Astronomy and Astrophysics of Maragha (RIAAM),
Maragha, Iran}

\begin{abstract}
In this paper, we consider a linearly charged dilatonic black holes and
study their thermodynamical behavior in the context of phase transition and
thermodynamic geometry. We show that, depending on the values of the
parameters, these type of black holes may enjoy two types of phase
transition. We also find that there are three critical behaviors near the
critical points for these black holes; nonphysical unstable to physical
stable, large to small, and small to large black holes phase transition.
Next, we employ a thermodynamical metric for studying thermodynamical
geometry of these black holes. We show that the characteristic behavioral of
Ricci scalar of this metric enables one to recognize the type of phase
transition and critical behavior of the black holes near phase transition
points. Finally, we will extend thermodynamical space by considering dilaton
parameter as extensive parameter. We will show that by this consideration,
Weinhold, Ruppeiner and Quevedo metrics provide extra divergencies which are
not related to any phase transition point whereas our new method is
providing an effective machinery.
\end{abstract}

\maketitle

\section{Introduction}

Recent astronomical observations indicate that our Universe is currently
undergoing a phase of accelerated expansion \cite{Perlmutter}. In the
context of standard cosmology, based on Einstein gravity, this acceleration
cannot be explained unless an unknown energy component usually dubbed ``dark
energy'' is proposed. Another way for explanation of such an acceleration is
the modification of the Einstein theory of gravity. In this regards, various
modifications of Einstein gravity proposed in the literatures. Among them
are Lovelock gravity \cite{Lovelock}, braneworld scenario \cite{Brax},
scalar-tensor theories \cite{Jordan,Giddings}, $f(R)$ gravity \cite{F(R)
gravity}, etc.

The studies on the black hole as a thermodynamic system date back to the
work of Hawking and Bekenstein \cite{HB}. According to the black holes
thermodynamics, the geometrical quantities such as horizon area and surface
gravity are related to the thermodynamic quantities such as entropy and
temperature. The first law of black hole thermodynamics implies that the
entropy and the temperature together with the energy (mass) of the black
hole satisfy $dE=TdS$ \cite{HB}. In recent years, the investigations on the
thermodynamical properties of the black holes have got a lot of interests.
In particular, thermodynamic properties of black holes in anti de-Sitter
(adS) spaces are improved in an extended phase space in which the
cosmological constant and its conjugate variable are considered as
thermodynamic pressure and volume, respectively \cite{Henneaux,Sekiwa,Brown}%
. In addition, there has been some proposals to consider constants (such as
Born-Infeld nonlinearity parameter, Gauss-Bonnet parameter, Newton constant
and etc.) as thermodynamical variable which contributes to thermodynamical
behavior of the system \cite{ConsTher}. It was shown that considering these
constants as thermodynamical variables will enrich the phase structure of
the black holes and describe new and interesting phenomena such as Van der
Waals like liquid/gas behavior. In this paper, motivated by these reasons,
we will extend the phase structure of dilatonic black holes by considering
dilaton parameter as a thermodynamical extensive parameter.

Recently, there has been several attempts in studying the phase transition
in dynamical context. It was shown that the quasinormal modes of a perturbed
black hole near critical point, exhibits different behaviors. In Ref. \cite%
{Quasi1}, it was argued that due to existence of normal modes only for
massless BTZ black holes, there is a phase transition from non-rotating BTZ
black holes. On the other hand, the four dimensional topological black holes
with scalar hair present the signature of the phase transition in their
quasi normal modes \cite{Quasi2}. Some evidences regarding second order
phase transition of a topological black hole to hairy one are given in Ref.
\cite{Quasi3}. Furthermore, a dramatic change in the slopes of quasinormal
frequencies in small and large black holes near the critical point was
observed for four-dimensional Reissner-Nordstr\"{o}m-adS black holes \cite%
{Quasi4}.

One of the interesting aspects of the black hole thermodynamics is stability
of black holes. In order to a black hole to be in thermal stability, its
heat capacity must be positive. This approach of studying the stability is
in the context of the canonical ensemble. Also, studying the heat capacity
of a system provides a machinery to study the phase transitions of the black
holes. There are two kinds of phase transitions; in the first case, changing
in the signature of the heat capacity is denoted as a type of phase
transition and therefore, the roots of the heat capacity are phase
transition points. So, we call these phase transitions type one. Another
type of phase transition is related to divergencies of the heat capacity.
This kind of phase transition is called the type two phase transition.

On the other hand, applying thermodynamical geometry to investigate the
phase transition of black holes has gained lots of attentions. The
pioneering studies were done by Weinhold \cite{Weinhold} and Ruppeiner \cite%
{Ruppeiner}. Weinhold proposed a metric on the space of equilibrium state,
introduced as the second derivatives of internal energy with respect to
entropy and other extensive quantities. The metric that Ruppeiner introduced
was defined as the minus second derivatives of entropy with respect to the
internal energy and other extensive quantities. It is worth noting that the
Ruppeiner's metric was conformal to Weinhold's metric with the inverse of
the temperature as the conformal factor \cite{Salamon}. On the other hand,
neither Weinhold nor Ruppeiner metrics were invariant under Legendre
transformation. Then, Quevedo \cite{Quevedo} proposed an approach to obtain
a metric which is Legendre invariant in the space of equilibrium state. His
motivation was based on the fact that the standard thermodynamics is
invariant under Legendre transformation. The formalism of geometrical
thermodynamics (GTs) implies that phase transition occurs at points where
the thermodynamics curvature is singular. As a consequence the curvature can
be interpreted as a measure of thermodynamic interaction \cite{CaiC}.
Recently, it was shown that employing Quevedo metrics for studying
geometrothermodynamics may not always lead to effective machinery. In other
words, there may be mismatch between phase transition points and
divergencies of the Ricci scalar of the Quevedo's metric. To solve the
problem, a new metric was proposed which has consistent results with heat
capacity \cite{HPEM}.

The outline of this paper is as follow. In the next section, we shall review
charged dilaton black holes and their thermodynamic quantities. In section
III, we will introduce the approaches for studying phase transitions of
these black holes in the context of heat capacity and GTs and study the
stability of these black holes. Then, we continue our paper and investigate
the existence of the phase transitions in the context of two mentioned
approaches. We also introduce some concepts such as stability, phase
transition and GTs of black holes in this section. In section IV, we discuss
the obtained diagrams of the phase transition, heat capacity and GTs of the
dilaton black holes. In section V, we consider dilatonic parameter, $\alpha $
as an extensive parameter and investigate the effects of it on different
approaches of GTs. The last section is devoted to closing remarks.

\section{Charged Black hole solutions in dilaton gravity}

The ($n+1$)-dimensional action of Einstein-Maxwell-dilaton (EMd) gravity has
the following form \cite{Sheykhi2007}
\begin{equation}
{I}=\frac{1}{16\pi }\int d^{n+1}x\sqrt{-g}\left[ \mathcal{R}-\frac{4}{n-1}%
\left( \nabla \Phi \right) ^{2}-V\left( \Phi \right) -e^{-4\alpha \Phi
/\left( n-1\right) }F_{\mu \nu }F^{\mu \nu }\right] ,  \label{Action}
\end{equation}%
where $\mathcal{R}$ is the Ricci scalar curvature, $F_{\mu \nu }=\partial
_{\mu }A_{\nu }-\partial _{\nu }A_{\mu }\mathcal{\ }$is the electromagnetic
field tensor and $A_{\mu }$ is the electromagnetic potential, $\Phi $ is the
dilaton field and $V\left( \Phi \right) $ is a potential for $\Phi $. Here, $%
\alpha $ is a constant determining the strength of coupling of the scalar
and electromagnetic field. One can vary action (\ref{Action}) with respect
to the gravitational field $g_{\mu \nu }$, the dilaton field $\Phi $ and the
gauge field $A_{\mu }$, which leads to the following field equations,
respectively,
\begin{equation}
R_{\mu \nu }=\frac{4}{n-1}\left( \partial _{\mu }\Phi \partial _{\nu }\Phi +%
\frac{1}{4}g_{\mu \nu }V(\Phi )\right) +2e^{-4\alpha \Phi /\left( n-1\right)
}\left( F_{\mu \eta }F_{\nu }^{\eta }-\frac{1}{2\left( n-1\right) }g_{\mu
\nu }F_{\lambda \eta }F^{\lambda \eta }\right),  \label{dilaton equation(I)}
\end{equation}%
\begin{equation}
\nabla ^{2}\Phi =\frac{n-1}{8}\frac{\partial V}{\partial \Phi }-\frac{\alpha
}{2}e^{-4\alpha \Phi /\left( n-1\right) }F_{\lambda \eta }F^{\lambda \eta },
\label{dilaton equation(II)}
\end{equation}%
\begin{equation}
\nabla _{\mu }\left( e^{-4\alpha \Phi /\left( n-1\right) }F^{\mu \nu
}\right) =0.  \label{Maxwell equation}
\end{equation}

We consider an ($n+1$)-dimensional spacetime with the following line element
\begin{equation}
ds^{2}=-f(r)dt^{2}+\frac{dr^{2}}{f(r)}+r^{2}R^{2}(r)h_{ij}dx^{i}dx^{j},
\label{metric}
\end{equation}%
where $f(r)$ and $R(r)$ are functions of $r$ which should be determined, and
$h_{ij}dx^{i}dx^{j}$ is the line element for an $(n-1)$-dimensional subspace
with $n(n-1)k$ constant curvature and volume $\omega _{n-1}$. We should note
that the constant $k$ indicates that the boundary of $t=\mathrm{constant}$
and $r=\mathrm{constant}$ can be a positive (elliptic), zero (flat) or
negative (hyperbolic) constant curvature hypersurface. The Maxwell equation (%
\ref{Maxwell equation}) can be integrated immediately to give \cite%
{Sheykhi2007}
\begin{equation}
F_{tr}=\frac{qe^{4\alpha \Phi /\left( n-1\right) }}{\left( rR\right) ^{n-1}}.
\label{Ftr eq}
\end{equation}

We consider the dilaton potential of the form \cite{Sheykhi2007}
\begin{equation}
V(\Phi )=\frac{k\left( n-1\right) \left( n-2\right) \alpha ^{2}}{b^{2}\left(
\alpha ^{2}-1\right) }e^{\frac{4\Phi }{\alpha \left( n-1\right) }}+2\Lambda
e^{\frac{4\alpha \Phi }{n-1}},  \label{V(Phi)}
\end{equation}
where $\Lambda$ is a free parameter which plays the role of the cosmological
constant. For later convenience, we redefine it as $\Lambda =-n\left(
n-1\right) /2l^{2}$, where $l$ is a constant with dimension of length. It is
notable that, this kind of potential was previously investigated by a number
of authors both in the context of Friedman-Robertson-Walker (FRW) scalar
field cosmologies \cite{Ozer} and EMd black holes \cite{Chan}.

On the other hand, in order to solve the system of equations (\ref{dilaton
equation(I)}) and (\ref{dilaton equation(II)}) for three unknown functions $%
f(r)$, $R(r)$ and $\Phi (r)$, we make the ansatz \cite{Dehghani}
\begin{equation}
R(r)=e^{2\alpha \Phi /\left( n-1\right) }.  \label{R(r)}
\end{equation}

Using Eqs. (\ref{Ftr eq})--(\ref{R(r)}) and the metric (\ref{metric}), one
can easily show that equations (\ref{dilaton equation(I)}) and (\ref{dilaton
equation(II)}) have solutions of the form \cite{Sheykhi2007}
\begin{eqnarray}
f(r) &=&-\frac{k\left( n-2\right) \left( \alpha ^{2}+1\right) ^{2}}{\left(
\alpha ^{2}-1\right) \left( \alpha ^{2}+n-2\right) }\left( \frac{b}{r}%
\right) ^{-2\gamma }-\frac{m}{r^{\left( n-1\right) \left( 1-\gamma \right)
-1}}+\frac{2\Lambda \left( \alpha ^{2}+1\right) ^{2}r^{2}}{\left( n-1\right)
\left( \alpha ^{2}-n\right) }\left( \frac{b}{r}\right) ^{2\gamma }  \nonumber
\\
&&+\frac{2q^{2}\left( \alpha ^{2}+1\right) ^{2}}{\left( n-1\right) \left(
\alpha ^{2}+n-2\right) r^{2\left( n-2\right) }}\left( \frac{b}{r}\right)
^{-2\left( n-2\right) \gamma },  \label{f(r)}
\end{eqnarray}%
\begin{equation}
\Phi (r)=\frac{n-1}{2\left( \alpha ^{2}+1\right) }\ln \left( \frac{b}{r}%
\right) ,  \label{Phi(r)}
\end{equation}%
where $b$ is an arbitrary constant and $\gamma =\alpha ^{2}/\left( \alpha
^{2}+1\right) $. In the above expression, $m$ and $q$ are integration
constants which are related to the total mass and electric charge of the
black hole, respectively.

It is notable that, in the absence of a non-trivial dilaton ($\alpha =\gamma
=0$), the solution (\ref{f(r)}) reduces to
\begin{equation}
f(r)=k-\frac{m}{r^{n-2}}-\frac{2\Lambda }{n\left( n-1\right) }r^{2}+\frac{%
2q^{2}}{\left( n-1\right) \left( n-2\right) r^{2\left( n-2\right) }},
\end{equation}%
which describes an ($n+1$)-dimensional asymptotically adS topological black
hole with a positive, zero or negative constant curvature hypersurface \cite%
{Brill}.

\subsection{Thermodynamic quantities of EMd black holes}

Since our aim in the next section is to study GTs and heat capacity of EMd
black holes, therefore, here we give a brief review regarding the
calculation of the conserved charges and thermodynamic quantities in which
they have been obtained in \cite{Sheykhi2007}. It was shown that by using
the Hamiltonian approach, one can find the mass $M$ and charge $Q$ of the
EMd black holes as \cite{Sheykhi2007}
\begin{equation}
M=\frac{\left( n-1\right) b^{\left( n-1\right) \gamma }\omega _{n-1}}{16\pi
\left( \alpha ^{2}+1\right) }m,  \label{Mass}
\end{equation}%
\begin{equation}
Q=\frac{\omega _{n-1}}{4\pi }q.  \label{Charge}
\end{equation}

The Hawking temperature calculated for the topological EMd black hole on the
outer horizon $r_{+}$ has the following form \cite{Sheykhi2007}
\begin{eqnarray}
T &=&-\frac{k\left( n-2\right) \left( \alpha ^{2}+1\right) }{2\pi \left(
\alpha ^{2}+n-2\right) r_{+}}\left( \frac{b}{r_{+}}\right) ^{-2\gamma }-%
\frac{\left( n-\alpha ^{2}\right) m}{4\pi \left( \alpha ^{2}+1\right) }%
r_{+}^{\left( n-1\right) \left( 1-\gamma \right) }  \nonumber \\
&&-\frac{q\left( \alpha ^{2}+1\right) }{\pi \left( \alpha ^{2}+n-2\right)
r_{+}^{2n-3}}\left( \frac{b}{r_{+}}\right) ^{-2\left( n-2\right) \gamma },
\label{temperature(I)}
\end{eqnarray}%
where $r_{+}$ satisfy $f(r=r_{+})=0$. One can obtain the entropy of the
topological EMd black hole by employing the area law. One finds \cite%
{Sheykhi2007}
\begin{equation}
S=\frac{b^{\left( n-1\right) \gamma }\omega _{n-1}}{4}r_{+}^{\left(
n-1\right) \left( 1-\gamma \right) }.  \label{Entropy}
\end{equation}

\section{Thermal stability, phase transition and GTs}

In this section, first, we study stability and phase transition of the EMd
black holes by considering the heat capacity of the solutions. Next, we
consider the GTs approach for studying phase transitions. We investigate the
effects of dilaton parameter and compare the results of both approaches.

There are several approaches for studying stability of the black holes. One
of these approaches is related to studying the perturbed black holes and see
if and how they acquire stable state and will be in equilibrium. This
approach is known as dynamical stability of black holes. In this paper, we
are not interested in dynamical stability of black holes but thermal one. We
focus our studies on thermal stability of charged black hole solutions in
the context of EMd theory through canonical ensemble. To do so, we calculate
the heat capacity and study its behavior.

Black holes should have a positive heat capacity in order to be thermally
stable. In other words, the positivity of the heat capacity ensures the
local thermal stability of the black holes. One can use the following
relation for the heat capacity
\begin{equation}
C_{Q}=T \left( \frac{\partial ^{2}M}{\partial S^{2}}\right)^{-1}_{Q}=T \left(%
\frac{\partial S}{\partial T}\right)_{Q}=T \left( \frac{\partial S}{\partial
r_{+}}\right)_{Q} \left( \frac{\partial T}{\partial r_{+}}\right)^{-1}_{Q} .
\label{CQ}
\end{equation}

On the other hand, one can use heat capacity for studying phase transitions
of black holes. In context of black holes, it is argued that roots of the
heat capacity are representing a type of phase transition. We call it type
one phase transition. The reasoning for such an assumption is due to fact
that system in case of this phase transition has a changing in signature of
the heat capacity. In other words, unstable system with negative heat
capacity goes under a phase transition and its heat capacity changes from
negative to positive. In addition, it is believed that the divergencies of
the heat capacity are also representing phase transitions of black holes. We
call these phase transitions, type two phase transitions. Therefore, the
phase transition points of the black holes in context of the heat capacity
are calculated with following relations
\begin{equation}
\left\{
\begin{array}{cc}
T=\left( \frac{\partial M}{\partial S}\right) _{Q}=0, & Type\ \ one \\
&  \\
\left( \frac{\partial ^{2}M}{\partial S^{2}}\right) _{Q}=0, & Type\ \ two%
\end{array}%
\right. .  \label{phase}
\end{equation}

Another approach for studying the phase transition of the black holes is
through GTs. There are several metrics that one can employ in order to build
geometrical spacetime by thermodynamical quantities. The well known ones are
Ruppeiner, Weinhold and Quevedo. It was previously argued that these metrics
may not provide us a completely flawless machinery for study GTs of specific
types of black holes \cite{HPEM}. Recently, a new metric (HPEM metric) was
proposed in order to solve the problems that other metrics may confront \cite%
{HPEM}. The denominator of the Ricci scalar of HPEM metric only contains
numerator and denominator of the heat capacity. In other words, divergence
points of the Ricci scalar of HPEM metric coincide with both types of phase
transitions of the heat capacity. The metric is in the following form
\begin{equation}
ds_{New}^{2}=\frac{SM_{S}}{\left( \Pi _{i=2}^{n}\frac{\partial ^{2}M}{%
\partial \chi _{i}^{2}}\right) ^{3}}\left(
-M_{SS}dS^{2}+\sum_{i=2}^{n}\left( \frac{\partial ^{2}M}{\partial \chi
_{i}^{2}}\right) d\chi _{i}^{2}\right) ,  \label{HPEM}
\end{equation}%
where $M_{S}=\partial M/\partial S$, $M_{SS}=\partial ^{2}M/\partial S^{2}$
and $\chi _{i}$ ($\chi _{i}\neq S$) are extensive parameters. In what
follows, we will study the stability and phase transition of the charged
dilatonic black holes in context of the heat capacity and GTs.

\section{Diagrams and Discussions}

In order to study the effects of different parameters on thermodynamical
behavior of the system we have plotted several diagrams in which the effects
of variations of $b$, $\alpha $ and $q$ are taken into account (Figs. \ref%
{Fig1} to \ref{Fig6} ).

\begin{figure}[tbp]
$%
\begin{array}{ccc}
\epsfxsize=6cm \epsffile{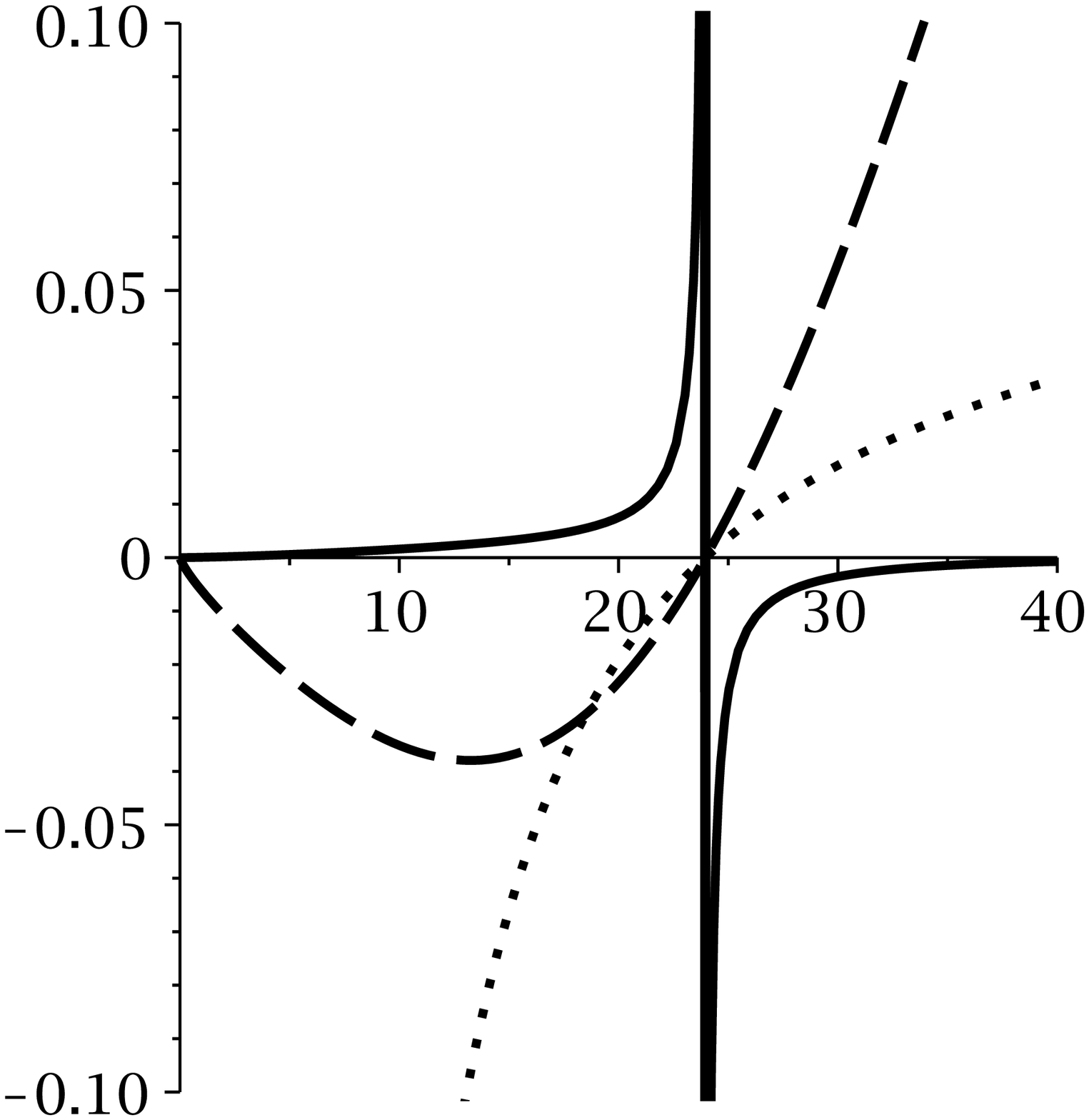} & \epsfxsize=6cm \epsffile{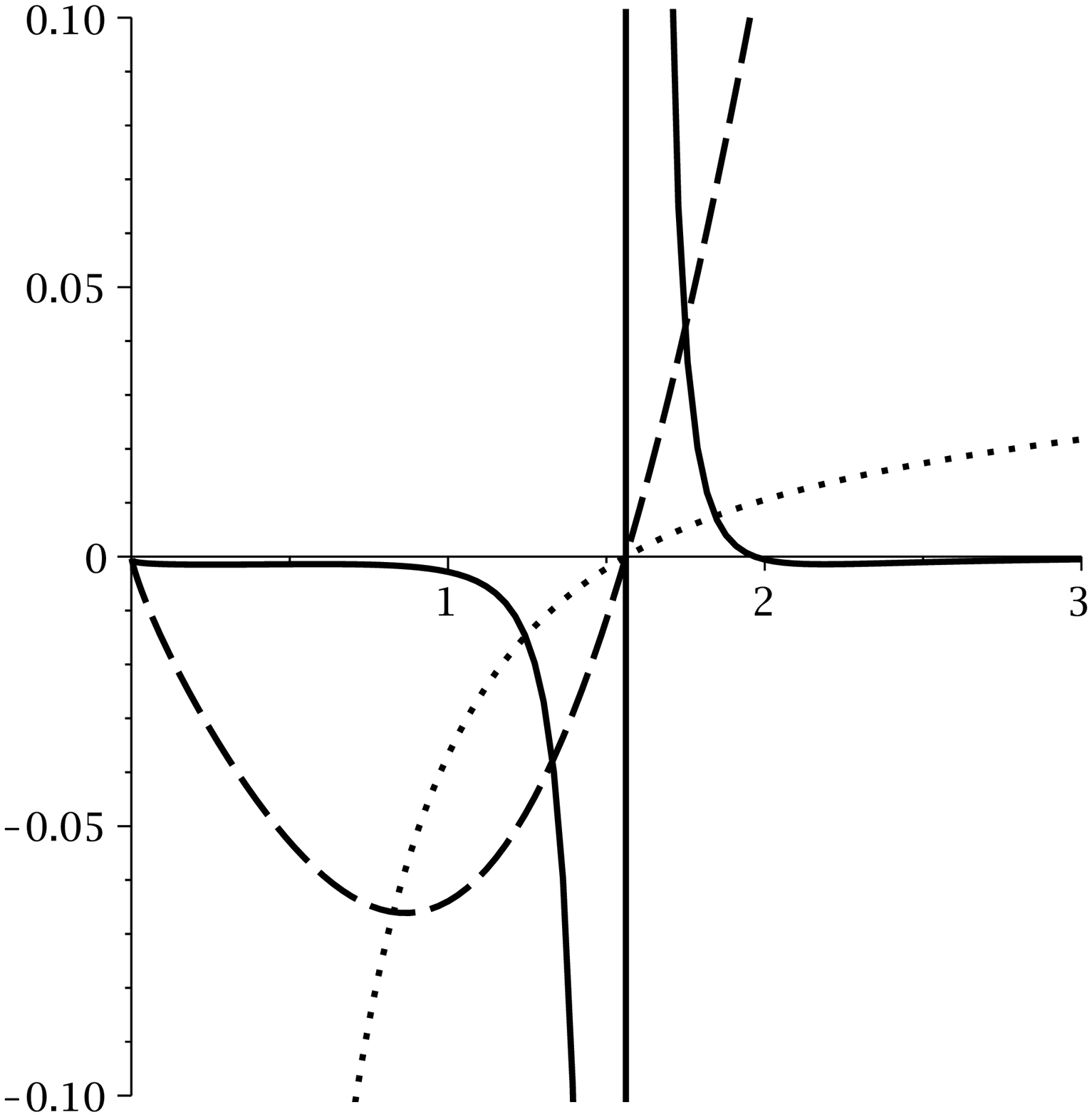} & %
\epsfxsize=6cm \epsffile{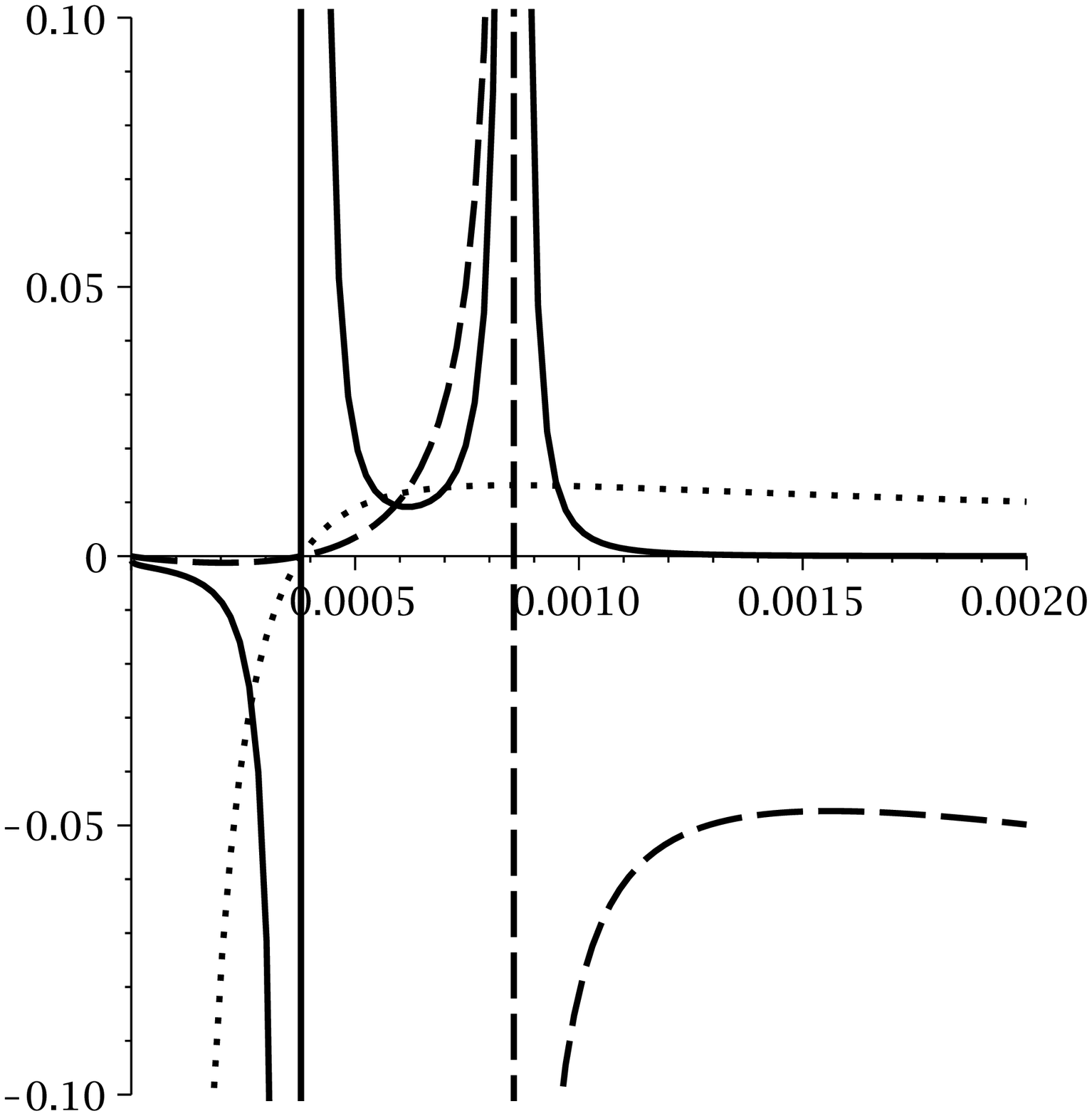}%
\end{array}
$%
\caption{$\mathcal{R}$ (continuous line), $C_{Q}$ (dashed line) and $T$
(dotted line) versus $r_{+}$ for $k=1$, $q=1$, $\protect\alpha=2$ and $n=5$
\newline
$b=0.05$ (left), $b=0.5$ (middle) and $b=5$ (right) }
\label{Fig1}
\end{figure}

\begin{figure}[tbp]
$%
\begin{array}{ccc}
\epsfxsize=6cm \epsffile{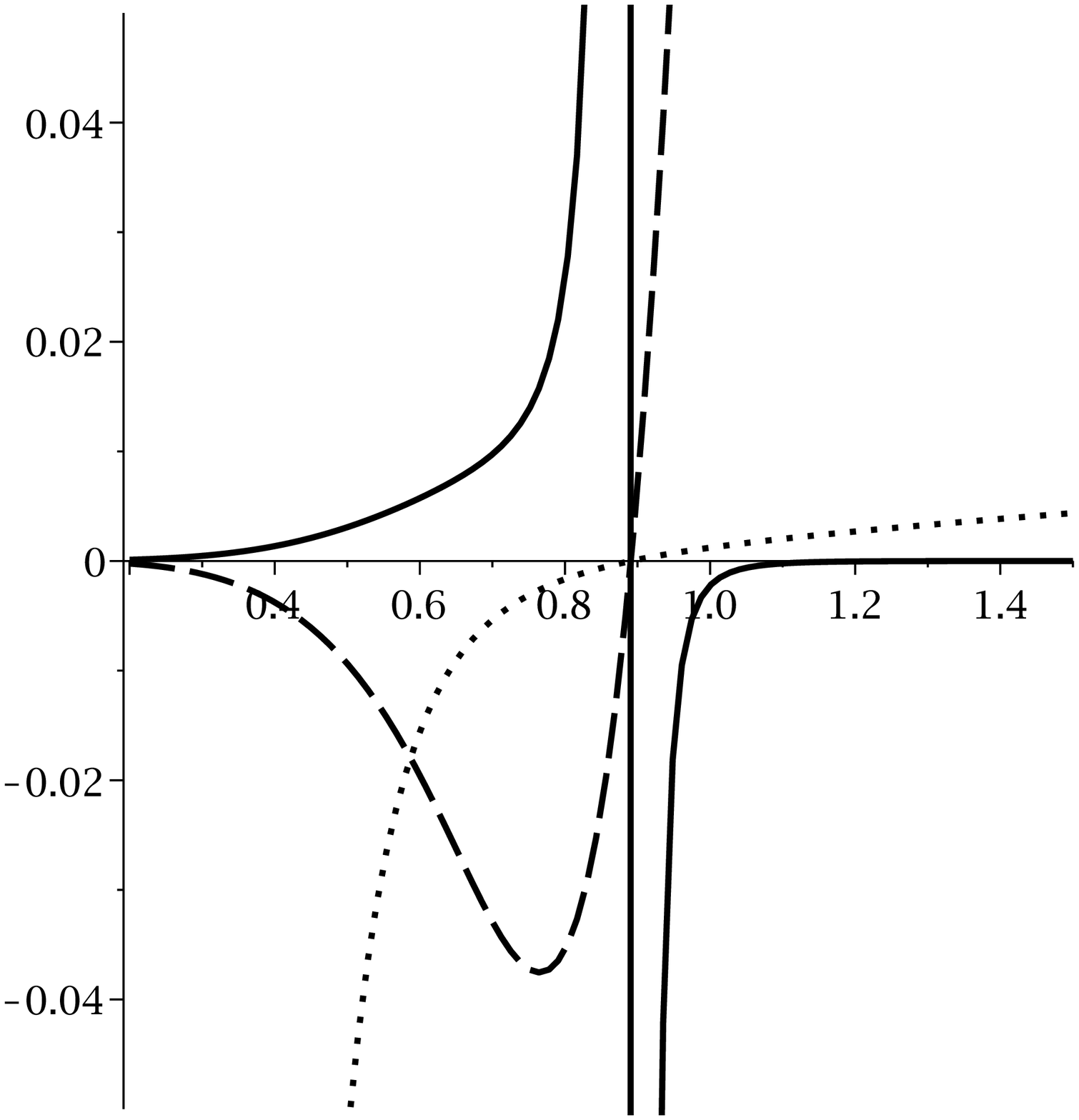} & \epsfxsize=6cm %
\epsffile{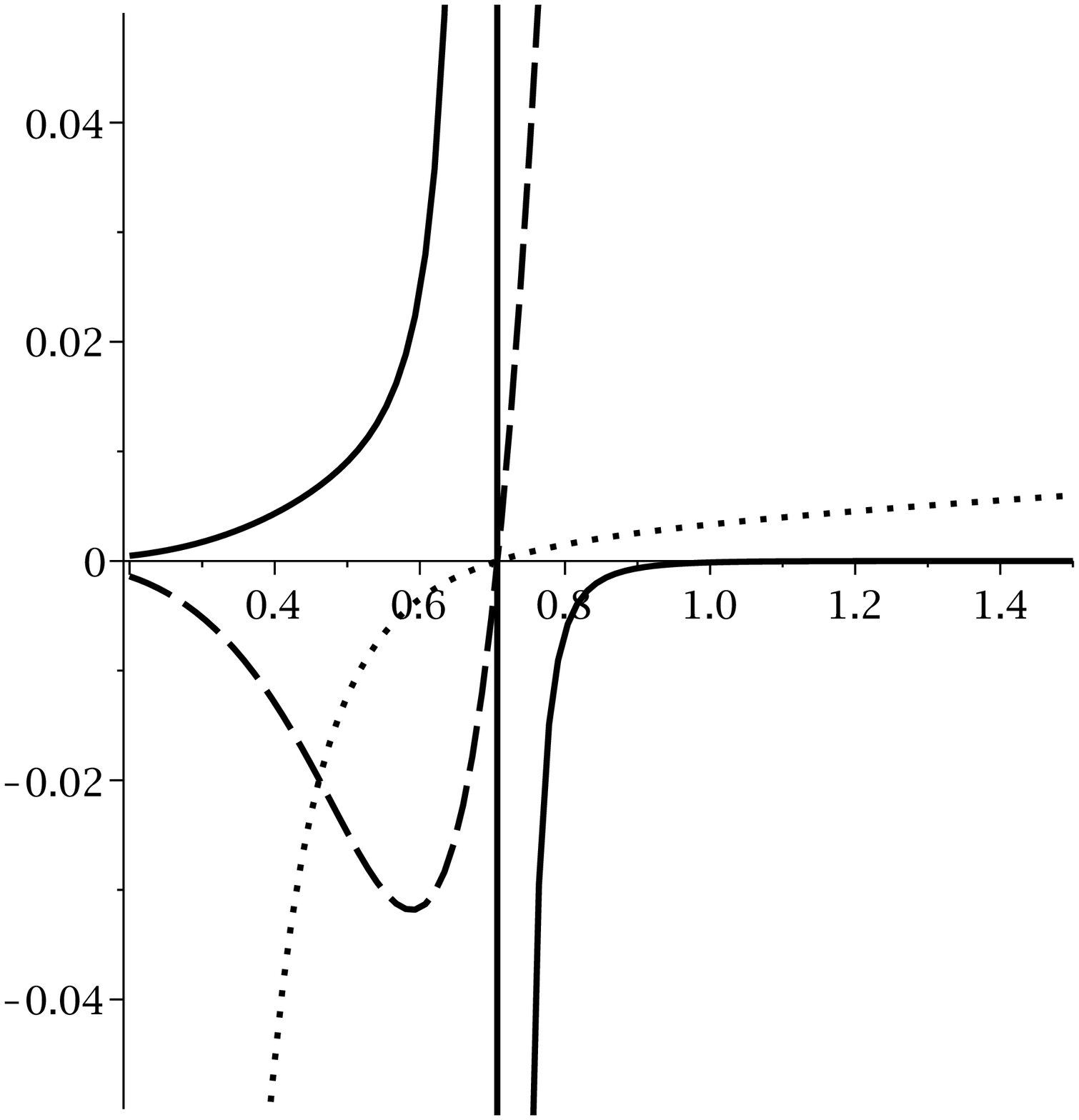} & \epsfxsize=6cm \epsffile{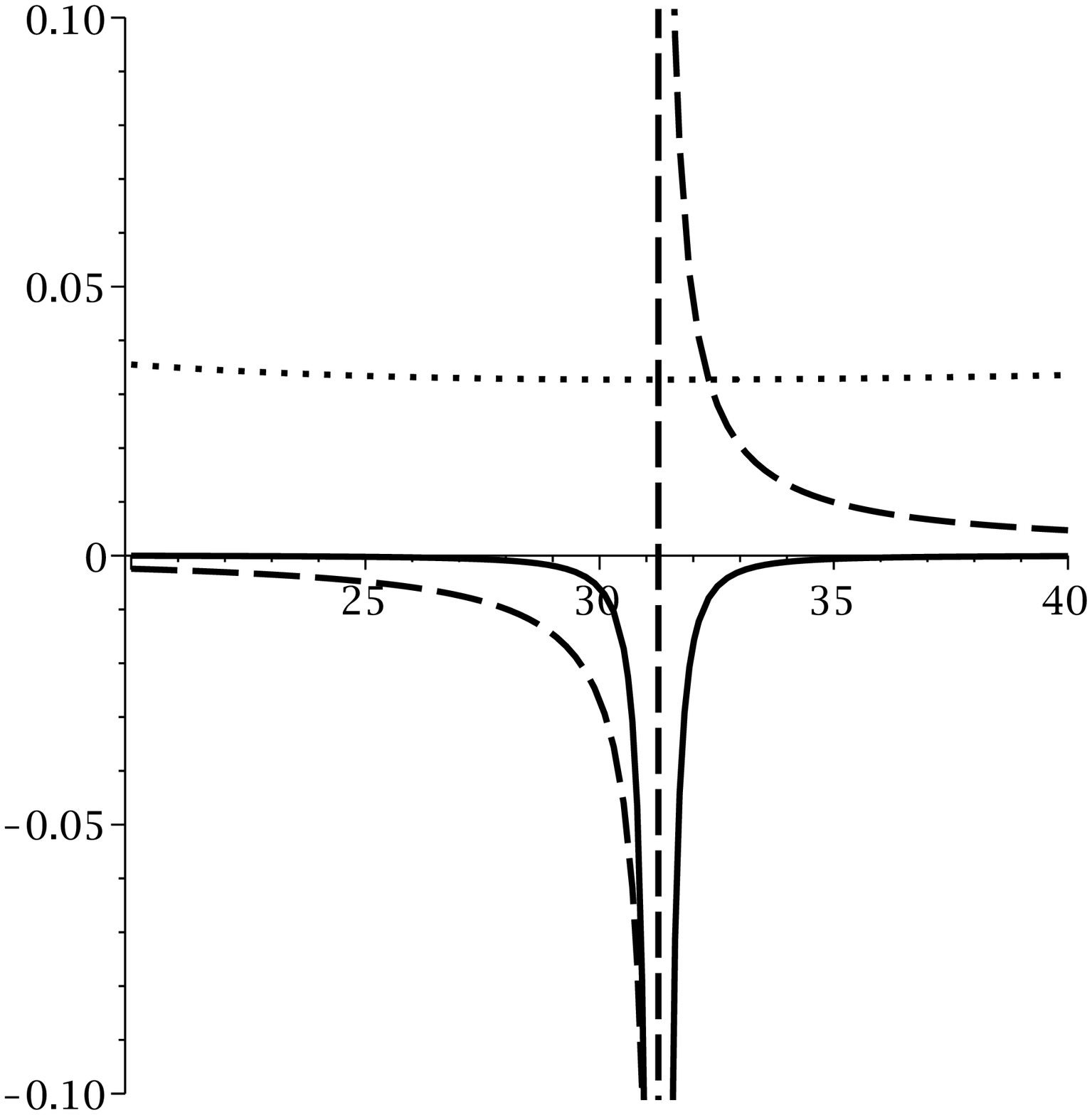}%
\end{array}
$%
\caption{$\mathcal{R}$ (continuous line), $C_{Q}$ (dashed line) and $T$
(dotted line) versus $r_{+}$ for $k=1$, $q=1$, $b=2$ and $n=5$ \newline
$\protect\alpha=0.05$ (left), $\protect\alpha=0.5$ (middle) and $\protect%
\alpha=5$ (right). }
\label{Fig2}
\end{figure}

\begin{figure}[tbp]
$%
\begin{array}{cc}
\epsfxsize=7cm \epsffile{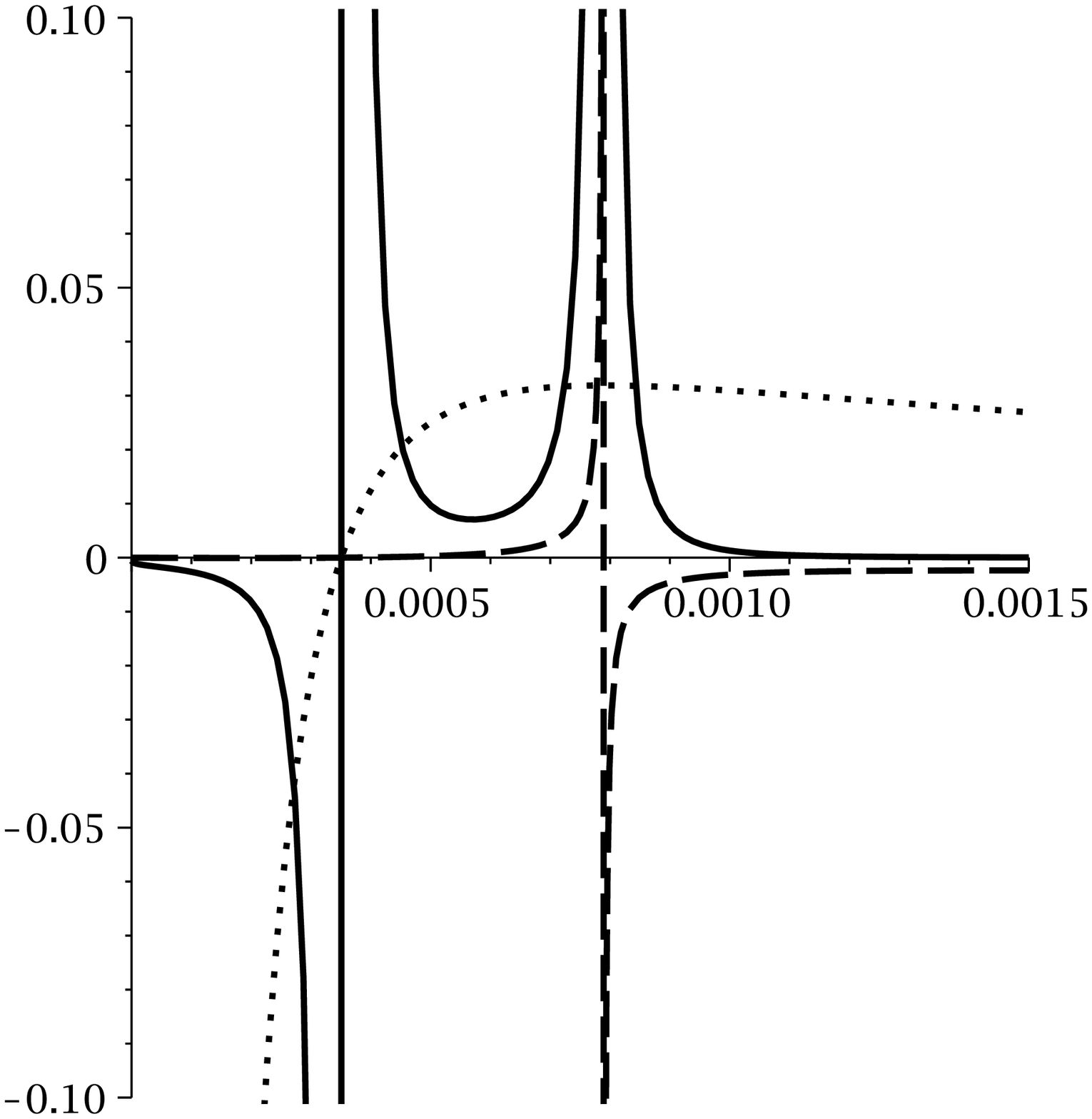} & \epsfxsize=7cm \epsffile{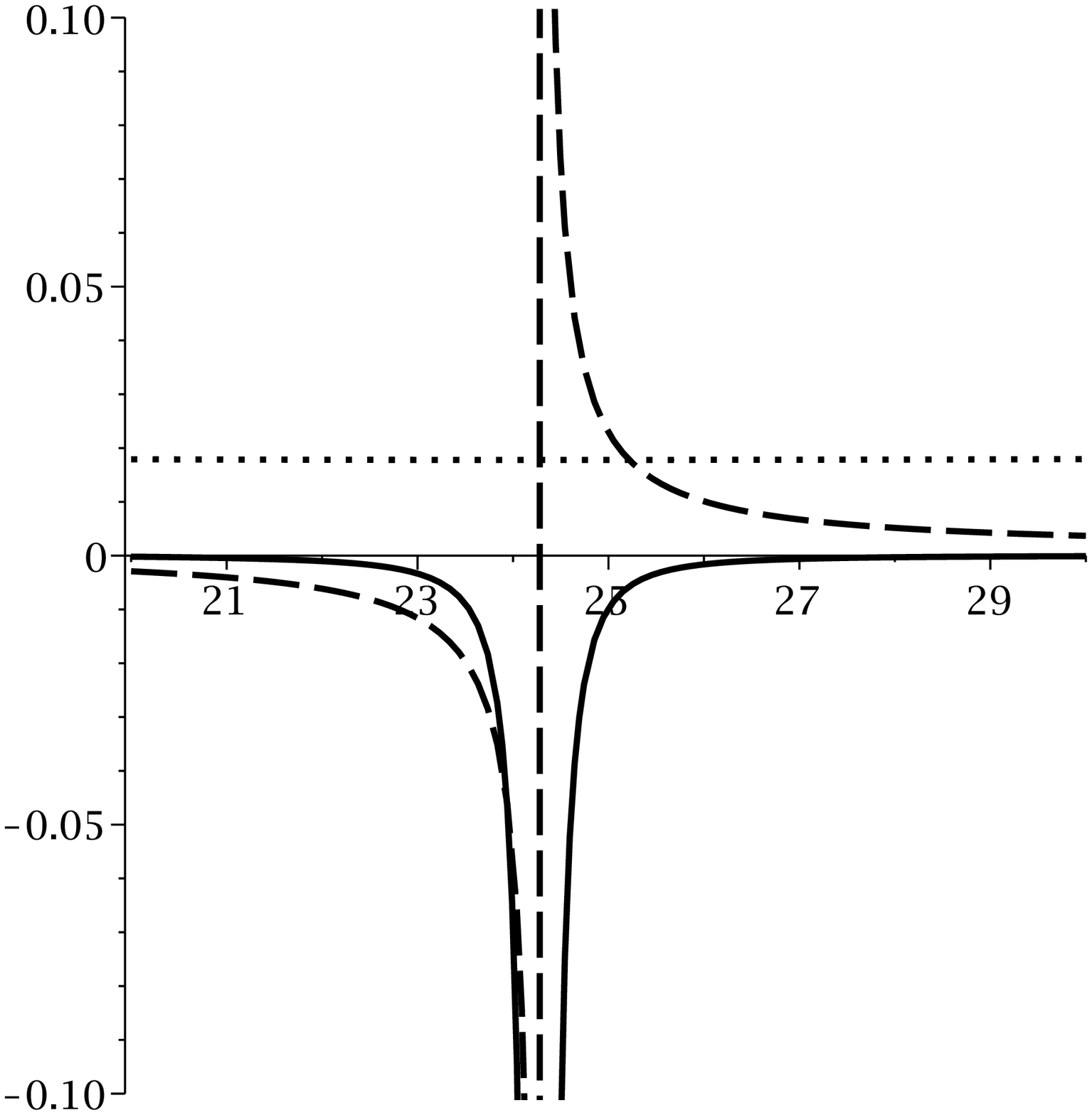}%
\end{array}
$%
\caption{$\mathcal{R}$ (continuous line), $C_{Q}$ (dashed line) and $T$
(dotted line) versus $r_{+}$ for $k=1$, $\protect\alpha=2$, $b=2$ and $n=5$
\newline
for different scales: $q=0.05$.}
\label{Fig3}
\end{figure}

\begin{figure}[tbp]
$%
\begin{array}{cc}
\epsfxsize=7cm \epsffile{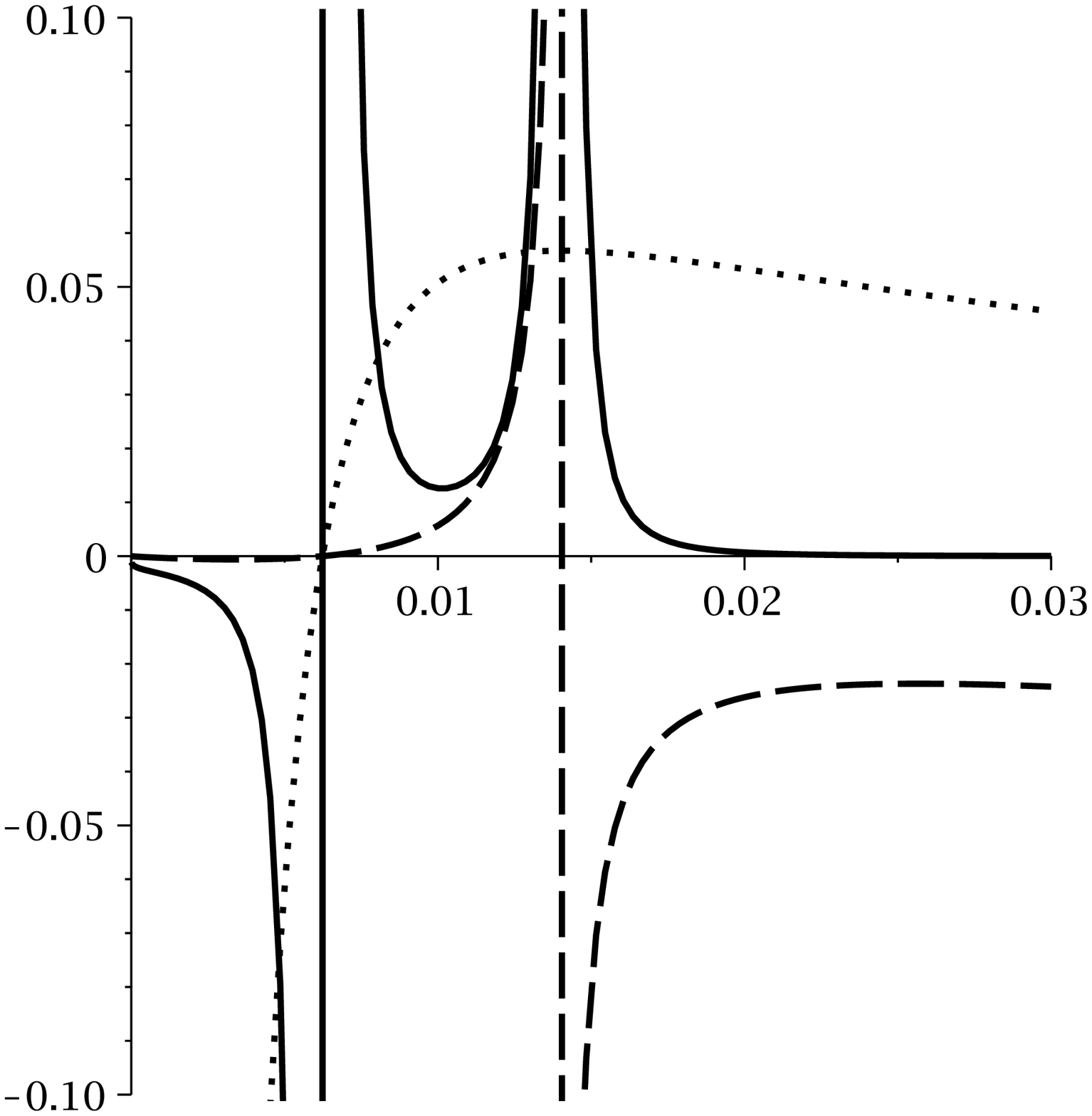} & \epsfxsize=7cm \epsffile{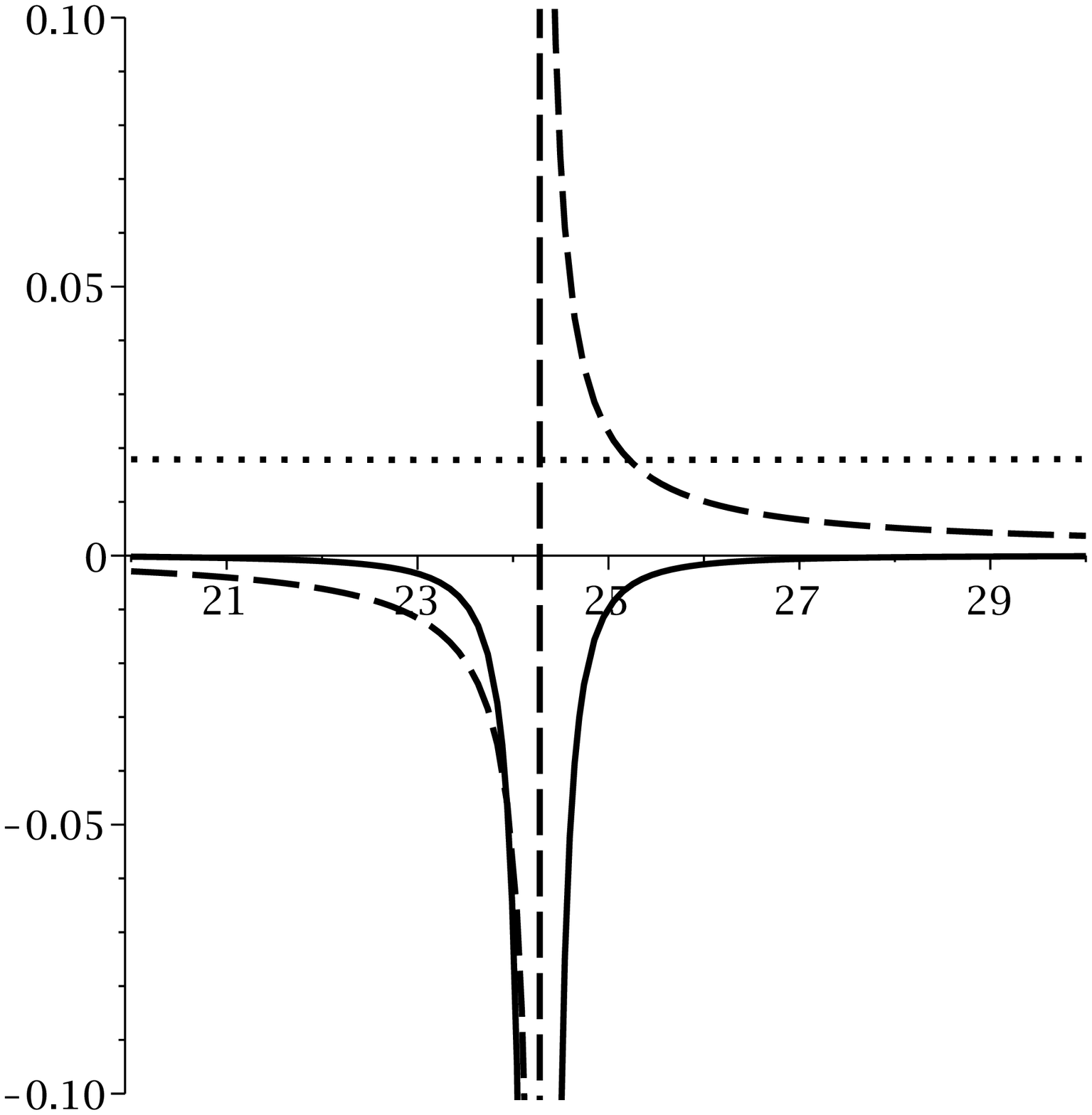}%
\end{array}
$%
\caption{$\mathcal{R}$ (continuous line), $C_{Q}$ (dashed line) and $T$
(dotted line) versus $r_{+}$ for $k=1$, $\protect\alpha=2$, $b=2$ and $n=5$
\newline
for different scales: $q=0.5$.}
\label{Fig4}
\end{figure}

\begin{figure}[tbp]
$%
\begin{array}{cc}
\epsfxsize=7cm \epsffile{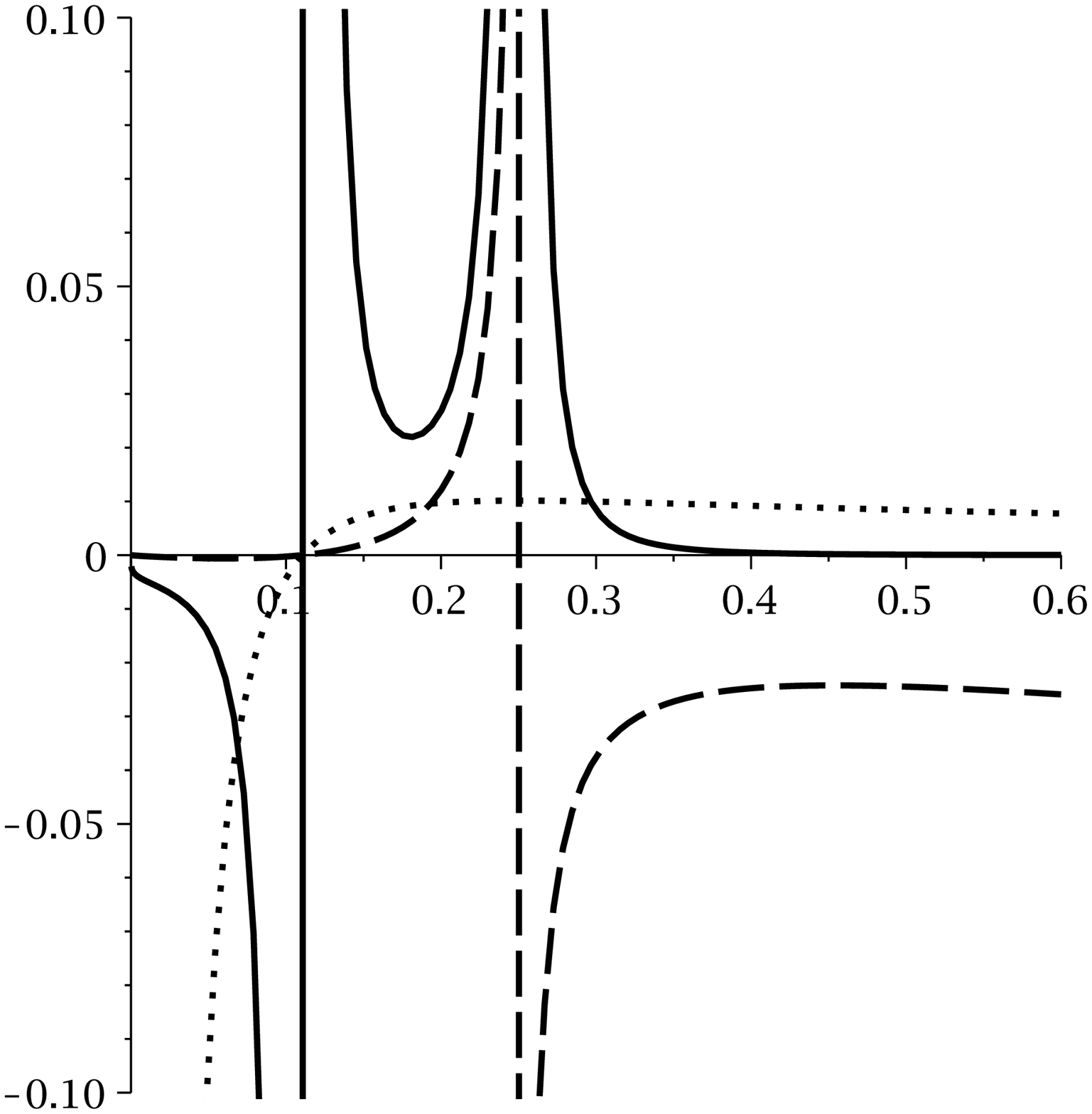} & \epsfxsize=7cm \epsffile{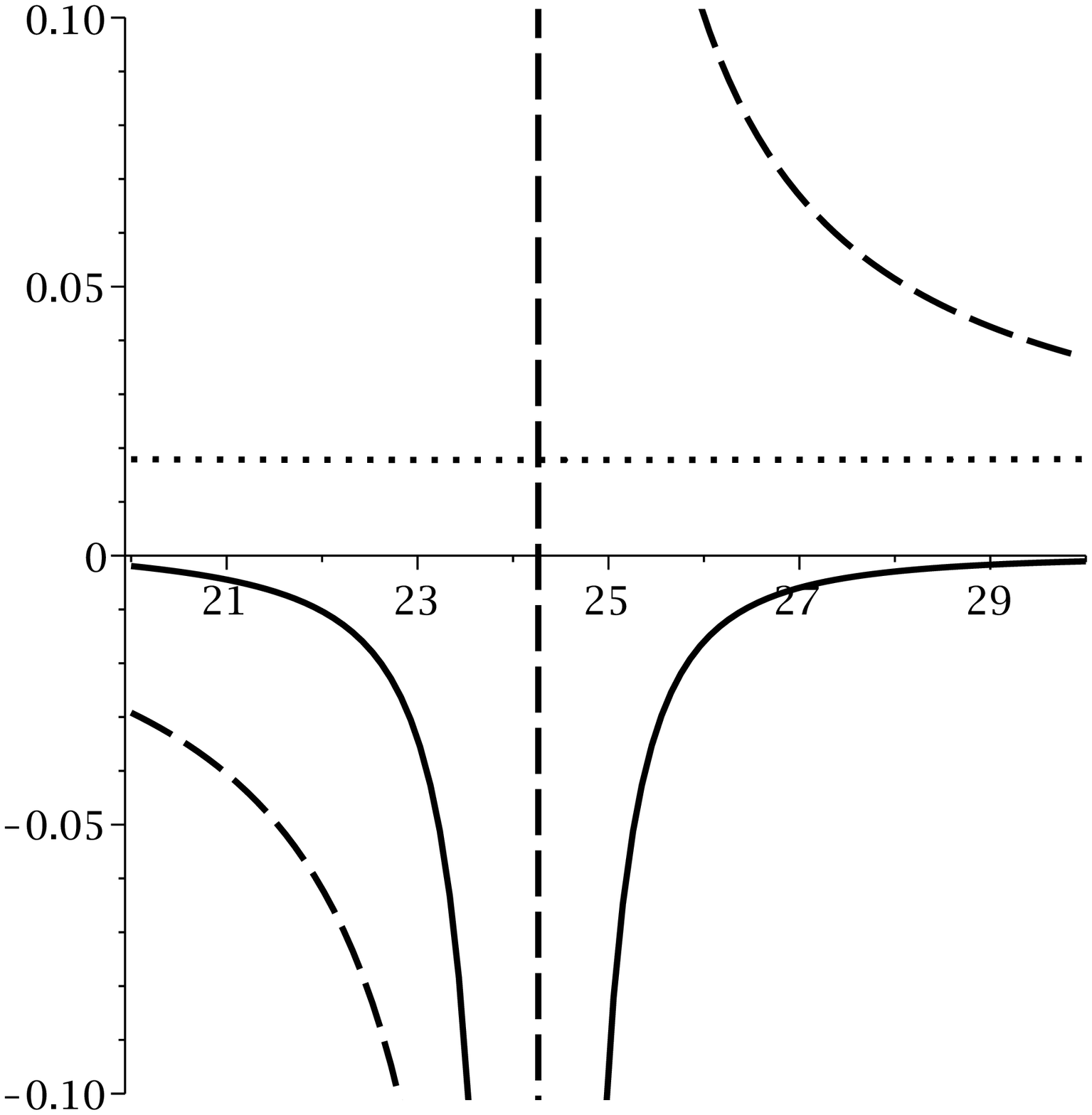}%
\end{array}
$%
\caption{$\mathcal{R}$ (continuous line), $C_{Q}$ (dashed line) and $T$
(dotted line) versus $r_{+}$ for $k=1$, $\protect\alpha=2$, $b=2$ and $n=5$
\newline
for different scales: $q=5$.}
\label{Fig5}
\end{figure}

\begin{figure}[tbp]
$%
\begin{array}{cc}
\epsfxsize=7cm \epsffile{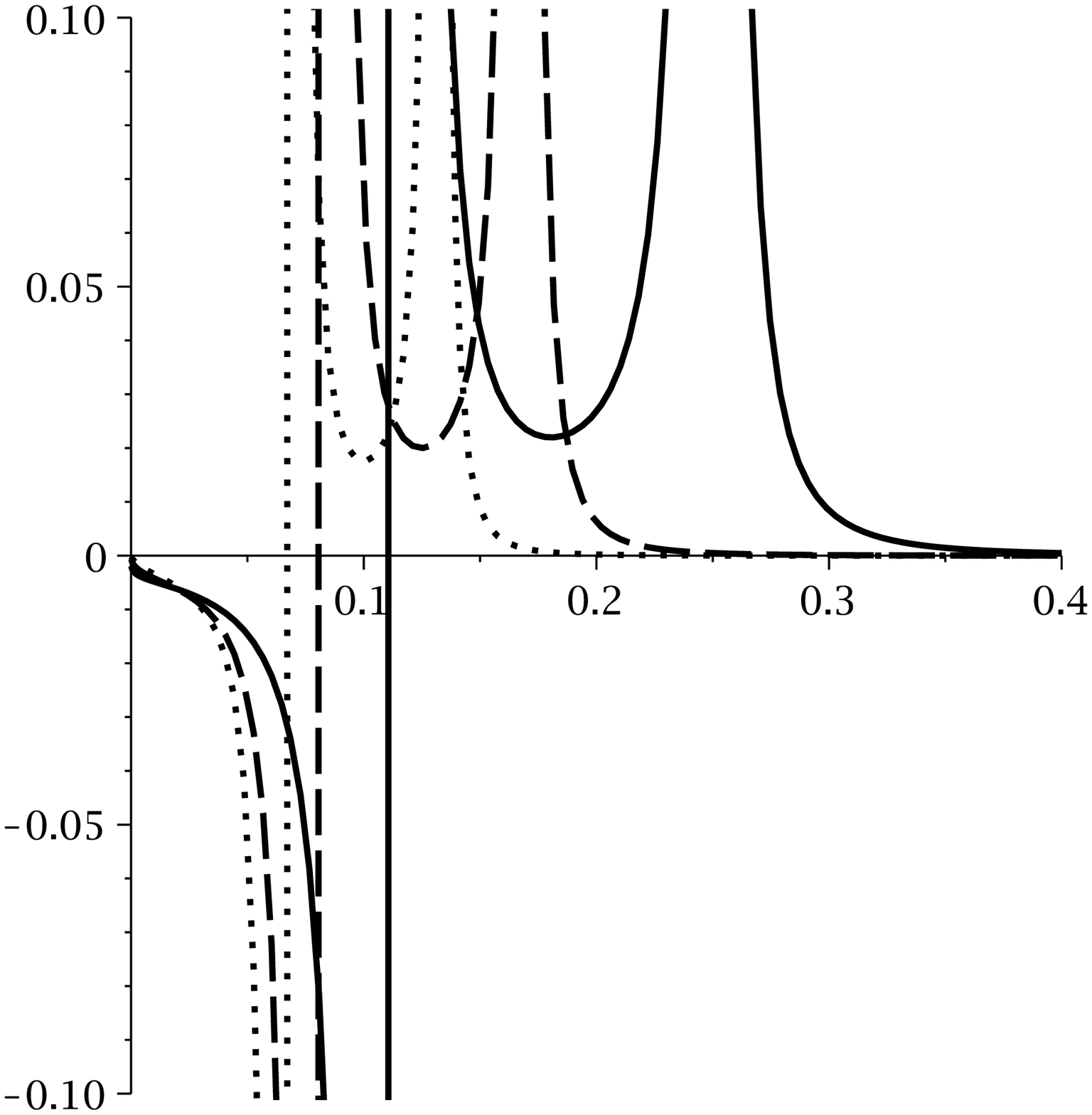} & \epsfxsize=7cm \epsffile{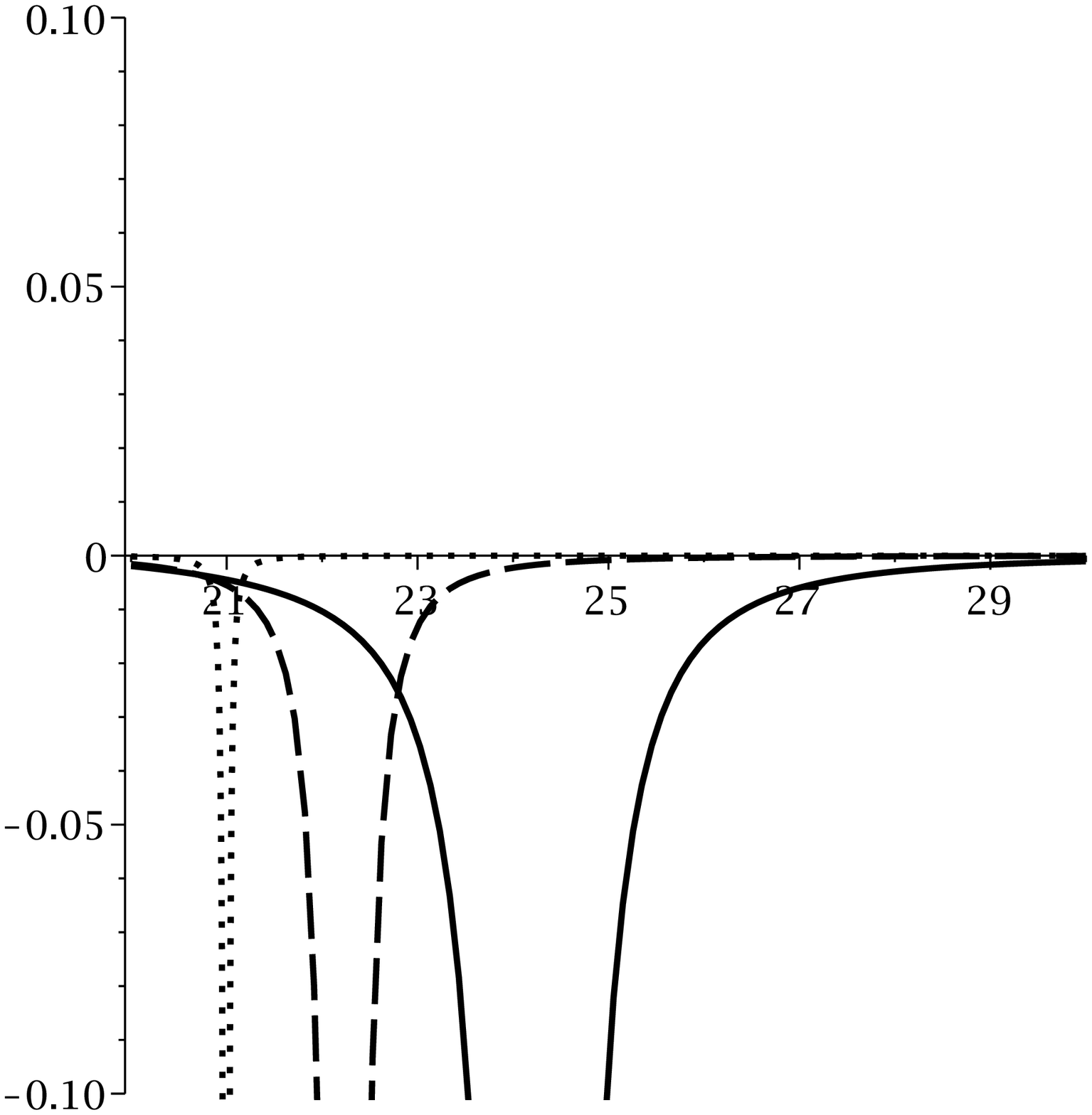}%
\end{array}
$%
\caption{$\mathcal{R}$ versus $r_{+}$ for $k=1$, $\protect\alpha=2$, $b=2$
and $q=5$ \newline
for different scales: $n=5$ (continuous line), $n=6$ (dashed line) and $n=7$
(dotted line) .}
\label{Fig6}
\end{figure}
First we will study the effects of the $b$. It is evident from plotted
graphs that for sufficiently small values of $b$, there is only type one
phase transition (Fig. \ref{Fig1} left and middle). In other words, the
denominator of the heat capacity is nonzero and there is only one root for
heat capacity. In place of this root, temperature and heat capacity changes
sign from negative to positive. Therefore, for the case of $r_{+}<r_{0}$ (in
which $C_{Q}\left( r_{+}=r_{0}\right) =0$), obtained solutions are
nonphysical unstable ones. It must be pointed out that, there is only one
divergency for Ricci scalar which coincides with root of the heat capacity.

For sufficiently large values of $b$ these black holes will enjoy one phase
transition type one and one phase transition type two (Fig. \ref{Fig1}
right). Therefore, the system will go under two type of phase transition. If
we denote the place of the divergency of the heat capacity with $r_{\infty }$%
, one can see that only for the region of $r_{0}<r_{+}<r_{\infty }$,
solutions are representing stable black holes. As for $r_{+}>r_{\infty }$,
system is unstable state. In this case too, the phase transition points of
the heat capacity and divergencies of the heat capacity coincide. It is
worthwhile to mention that both phase transition points of the heat capacity
are decreasing functions of $b$.

Next, we are taking the effects of variation of dilation parameter, $\alpha $
into account. Evidently, similar to the case of $b$, for small values of $%
\alpha $, there is only one phase transition of type one, which its place is
a decreasing function of $\alpha $ (Fig. \ref{Fig2} left and middle).
Interestingly, for sufficiently large values of $\alpha $, the phase
transition type one of the heat capacity will vanish and there will be only
one phase transition type two for these black holes (Fig. \ref{Fig2} right).
In this case for $r_{+}<r_{\infty }$, black holes are unstable whereas for $%
r_{+}=r_{\infty }$, they go under phase transition and acquire stable state.
The place of divergency of the heat capacity is an increasing function of $%
\alpha $. It is worthwhile to mention that in these cases too, Ricci scalar
of the considered thermodynamical metric has divergencies exactly in place
of phase transitions of heat capacity.

Our next effects of interest are the ones that are due to variation of
charge. As one can see, there are two phase transitions type two and one
phase transition type one for considered values for different parameters
(Figs. \ref{Fig3} to \ref{Fig5}). The root and smaller divergence point of
heat capacity are highly increasing functions of charge whereas the larger
divergence point of heat capacity is not effected considerably by variation
of charge.

Considering the number of the phase transition points, we have three phase
transition. In case of $r_{+}=r_{0}$, the temperature and heat capacity
changing sign from unstable nonphysical black holes to stable physical ones.
Next phase transition takes place in $r_{+}=r_{S}$ ($r_{S}$ is smaller
divergence point of heat capacity). Due to thermodynamical behavior (phase
transition from unstable state to stable one), there is a phase transition
from black holes with larger horizon radius to ones with smaller horizon
radius. In other words, in this place there is a phase transition of
larger/smaller black holes. In opposite, in case of phase transition that
takes place in $r_{+}=r_{L}$ ($r_{L}$ is larger divergence point of heat
capacity) there is a phase transition of smaller/larger black holes. In
other words, unstable black holes goes under phase transition and obtain
stable state with larger horizon radius.

Finally, as for dimensions, it is evident that root and divergence points of
heat capacity are decreasing functions of dimensionality (Fig. \ref{Fig6}).
It is worthwhile to mention that divergence points of Ricci scalar of
considered metric, in all the cases, coincide with phase transition points
of heat capacity. The behavior of the Ricci scalar near these divergence
points is different for each one of these phase transition points. In case
of root of heat capacity, there is a change in signature of the Ricci scalar
near corresponding divergency (Figs. \ref{Fig3} to \ref{Fig5} left). As for
phase transition of larger/smaller black holes, the signature of the Ricci
scalar is fixed and the divergency is toward $+\infty $ (Figs. \ref{Fig3} to %
\ref{Fig5} left) whereas in case of smaller/larger phase transition the
signature is fixed but it is toward $-\infty $ (Figs. \ref{Fig3} to \ref%
{Fig5} right). Therefore, one is able to recognize type of phase transition
and the behavior of it only by studying Ricci scalar of HPEM metric (Fig. %
\ref{Fig6}).

\section{Extending thermodynamical space}

In this section we would like to extend our phase space by considering the
dilaton-electromagnetic coupling parameter, $\alpha $, as an extensive
parameter. We shall explore the effects of this phase space extension in
different approaches of GTs. Therefore, our thermodynamical space will
change from $M(S,Q)$ to $M(S,Q,\alpha )$. This consideration causes the
Weinhold, Ruppeiner, Quevedo and HPEM metrics modified into following forms
\begin{equation}
ds^{2}=\left\{
\begin{array}{cc}
ds_{W}^{2}=Mg_{ab}^{W}dX^{a}dX^{b} & Weinhold \\
&  \\
ds_{R}^{2}=-T^{-1}Mg_{ab}^{R}dX^{a}dX^{b} & Ruppeiner \\
&  \\
\left( SM_{S}+QM_{Q}+\alpha M_{\alpha }\right) \left(
-M_{SS}dS^{2}+M_{QQ}dQ^{2}+M_{\alpha \alpha }d\alpha ^{2}\right) & Quevedo%
\text{ }Case\text{ }I \\
&  \\
SM_{S}\left( -M_{SS}dS^{2}+M_{QQ}dQ^{2}+M_{\alpha \alpha }d\alpha ^{2}\right)
& Quevedo\text{ }Case\text{ }II \\
&  \\
S\frac{M_{S}}{M_{QQ}^{3}M_{\alpha \alpha }^{3}}\left(
-M_{SS}dS^{2}+M_{QQ}dQ^{2}+M_{\alpha \alpha }d\alpha ^{2}\right) & HPEM%
\end{array}%
\right. .
\end{equation}

It is a matter of calculation to show that mentioned metric with these
specific structures have following denominators for their Ricci scalars
\begin{equation}
denom(\mathcal{R})=\left\{
\begin{array}{cc}
-2M^{3}\left( M_{Q\alpha }^{2}M_{SS}+M_{SQ}^{2}M_{\alpha \alpha }+M_{S\alpha
}^{2}M_{QQ}-M_{SS}M_{QQ}M_{\alpha \alpha }-2M_{SQ}M_{S\alpha }M_{Q\alpha
}\right) ^{2} & Weinhold \\
&  \\
-2M^{3}T^{3}\left( M_{Q\alpha }^{2}M_{SS}+M_{SQ}^{2}M_{\alpha \alpha
}+M_{S\alpha }^{2}M_{QQ}-M_{SS}M_{QQ}M_{\alpha \alpha }-2M_{SQ}M_{S\alpha
}M_{Q\alpha }\right) ^{2} & Ruppeiner \\
&  \\
2M_{SS}^{2}M_{QQ}^{2}M_{\alpha \alpha }^{2}\left( SM_{S}+QM_{Q}+\alpha
M_{\alpha }\right) ^{3} & Quevedo\;I \\
&  \\
2S^{3}M_{SS}^{2}M_{QQ}^{2}M_{\alpha \alpha }^{2}M_{S}^{3} & Quevedo\;II \\
&  \\
2S^{3}M_{SS}^{2}M_{S}^{3} & HPEM%
\end{array}%
\right. .
\end{equation}

\begin{figure}[tbp]
$%
\begin{array}{cc}
\epsfxsize=6cm \epsffile{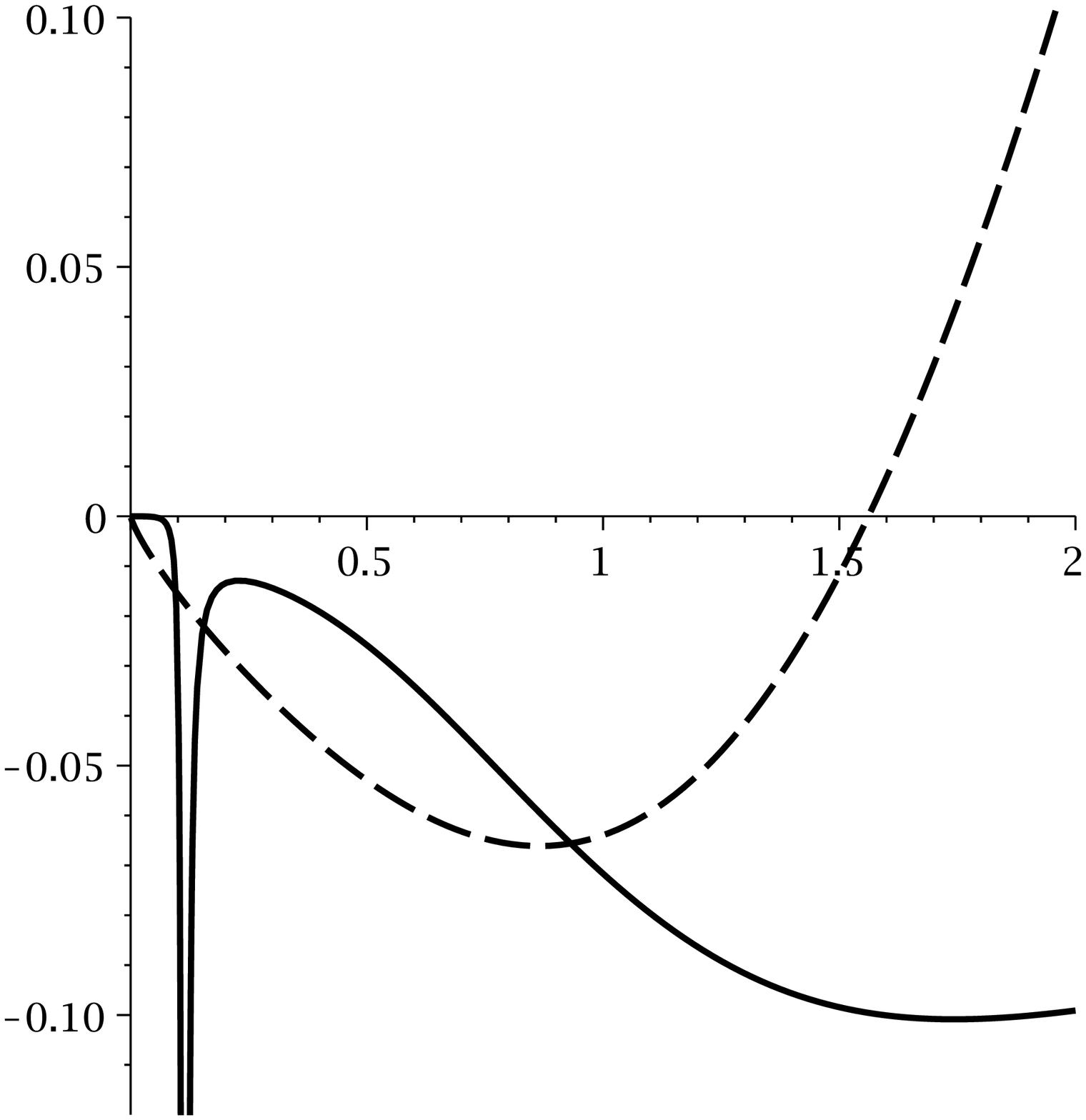} & \epsfxsize=6cm \epsffile{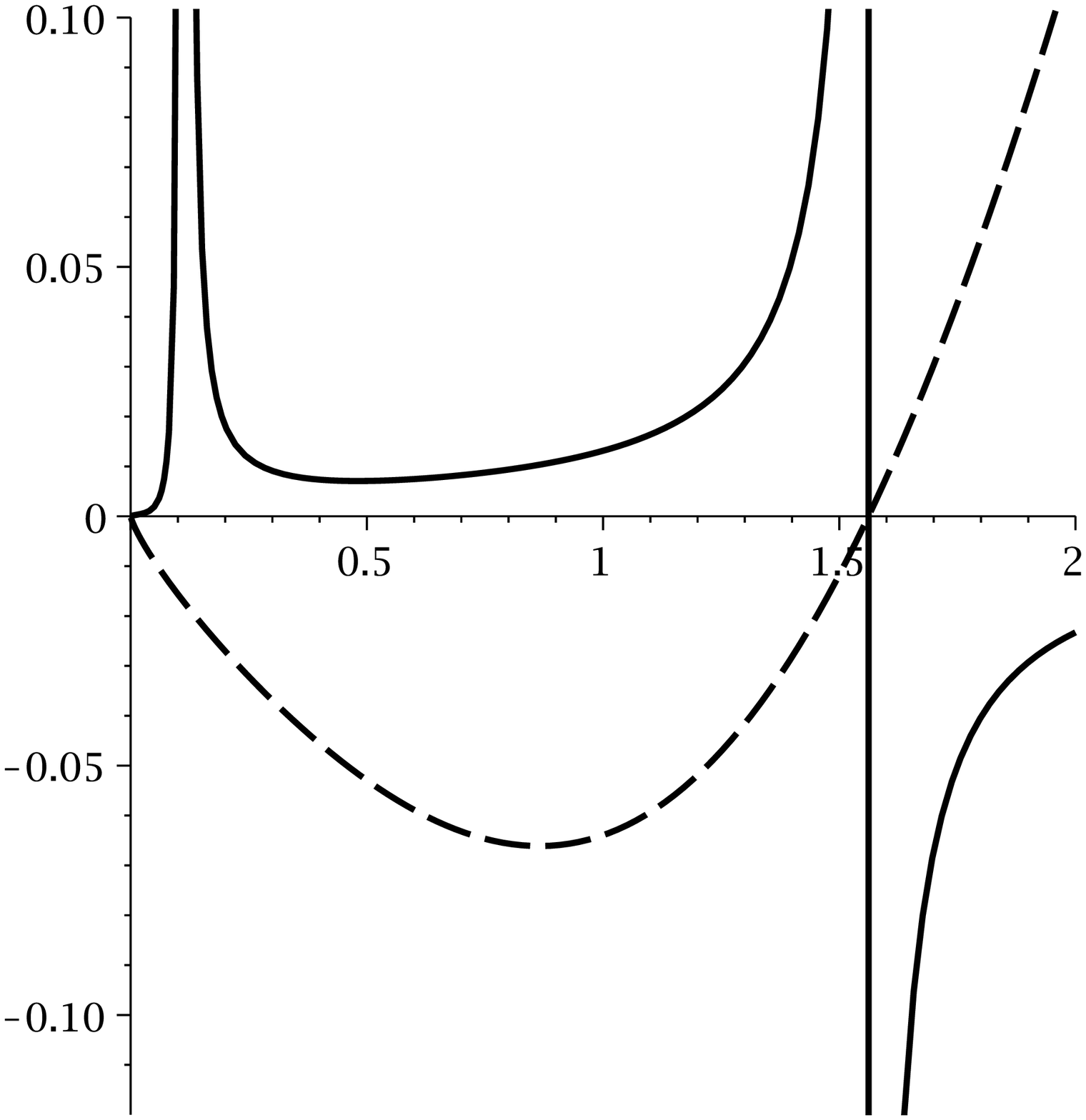}%
\end{array}
$%
\caption{$\mathcal{R}$ (continues line) and $C_{Q}$ (dashed line) versus $%
r_{+}$ for $l=1$, $\Lambda =-1$, $n=5$ and $q=1$, $\protect\alpha=2$, $b=0.5$
and $k=1$. \newline
left diagram: The Weinhold metric. \newline
right diagram: The Ruppeiner metric. }
\label{W}
\end{figure}

It is evident that, in general, for Weinhold case, due to obtained relation,
the coincidence of different types of the phase transition points and
divergencies of the Ricci scalar may not take place (Fig. \ref{W} left). As
for the Ruppeiner metric, the existence of $T$, in denominator of the Ricci
scalar ensures the coincidence of divergency of the Ricci scalar and phase
transition type one. As for the type two, due to other terms of the
denominator of the Ricci scalar, in general, the coincidence may not happen
or extra divergencies may be observed (Fig. \ref{W} right).

\begin{figure}[tbp]
$%
\begin{array}{cc}
\epsfxsize=6cm \epsffile{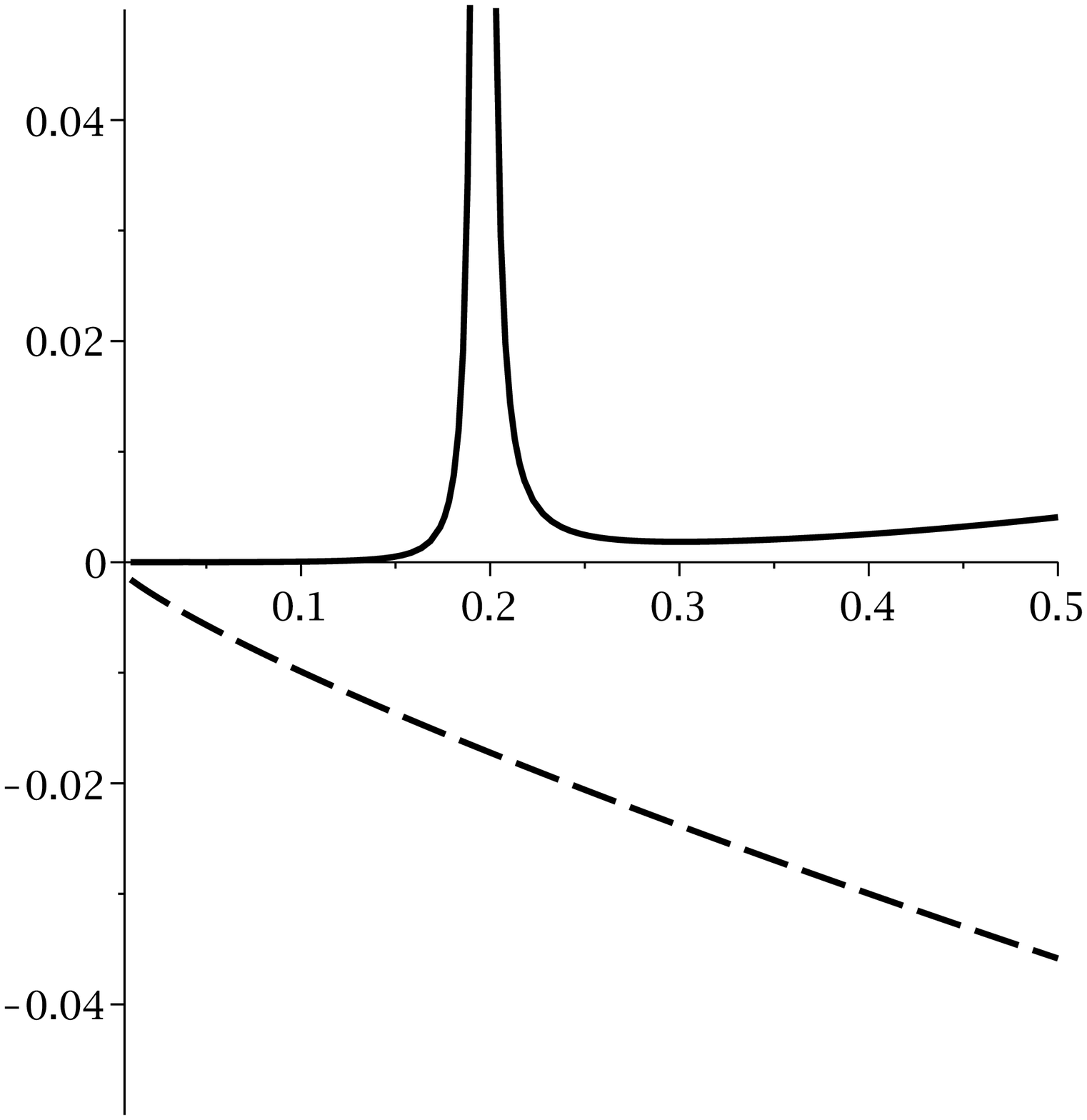} & \epsfxsize=6cm \epsffile{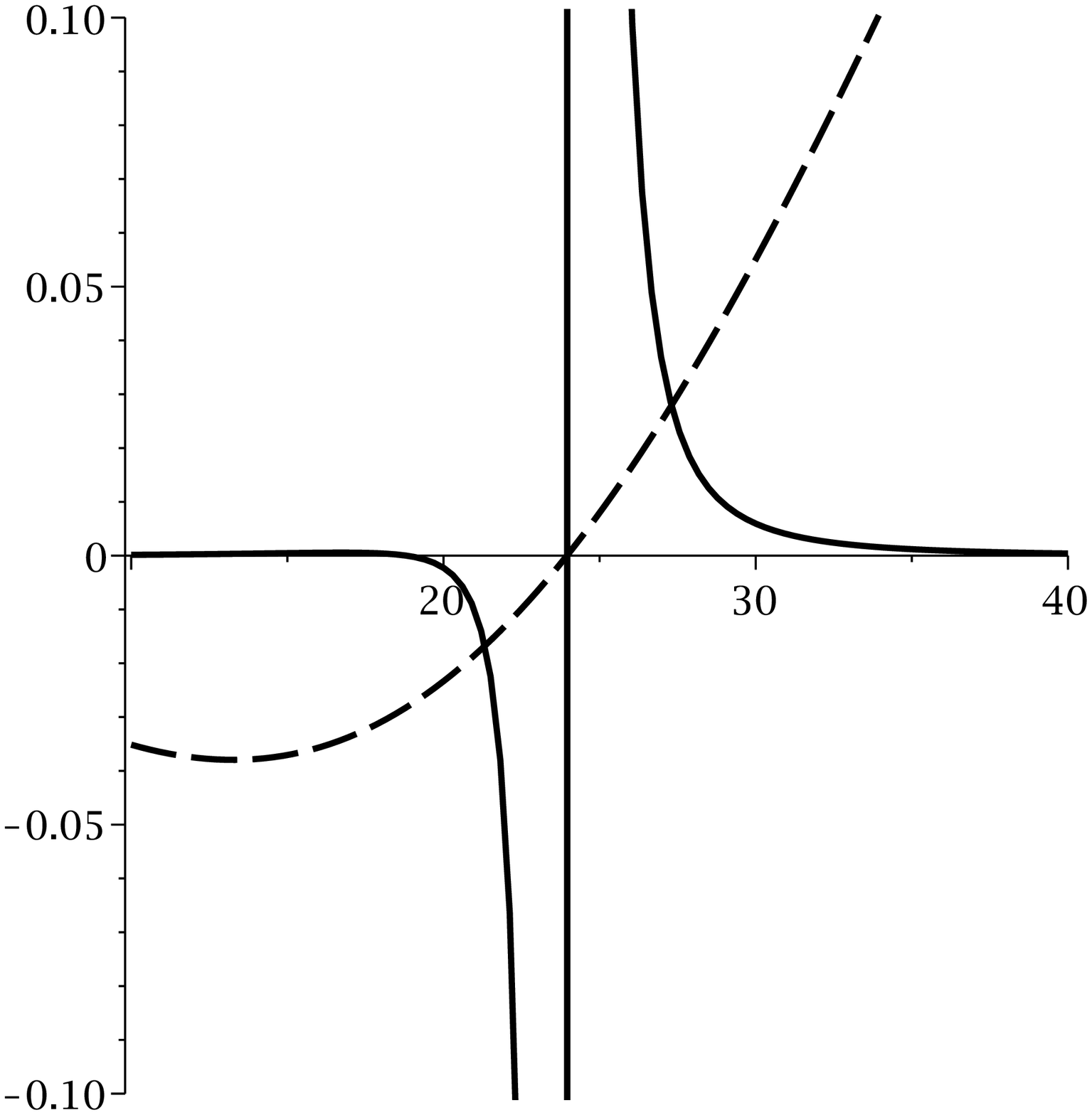} \\
\epsfxsize=6cm \epsffile{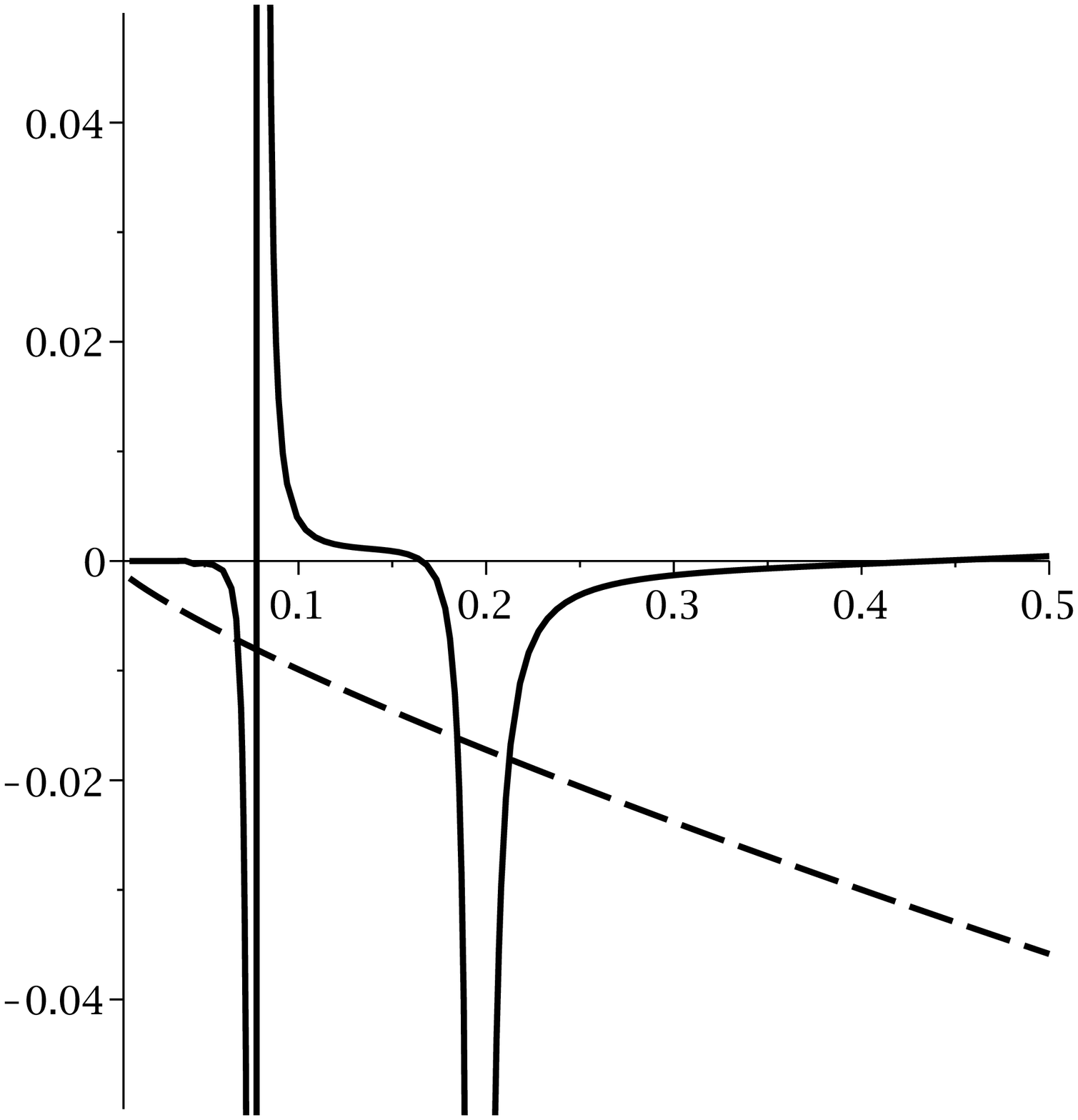} & \epsfxsize=6cm \epsffile{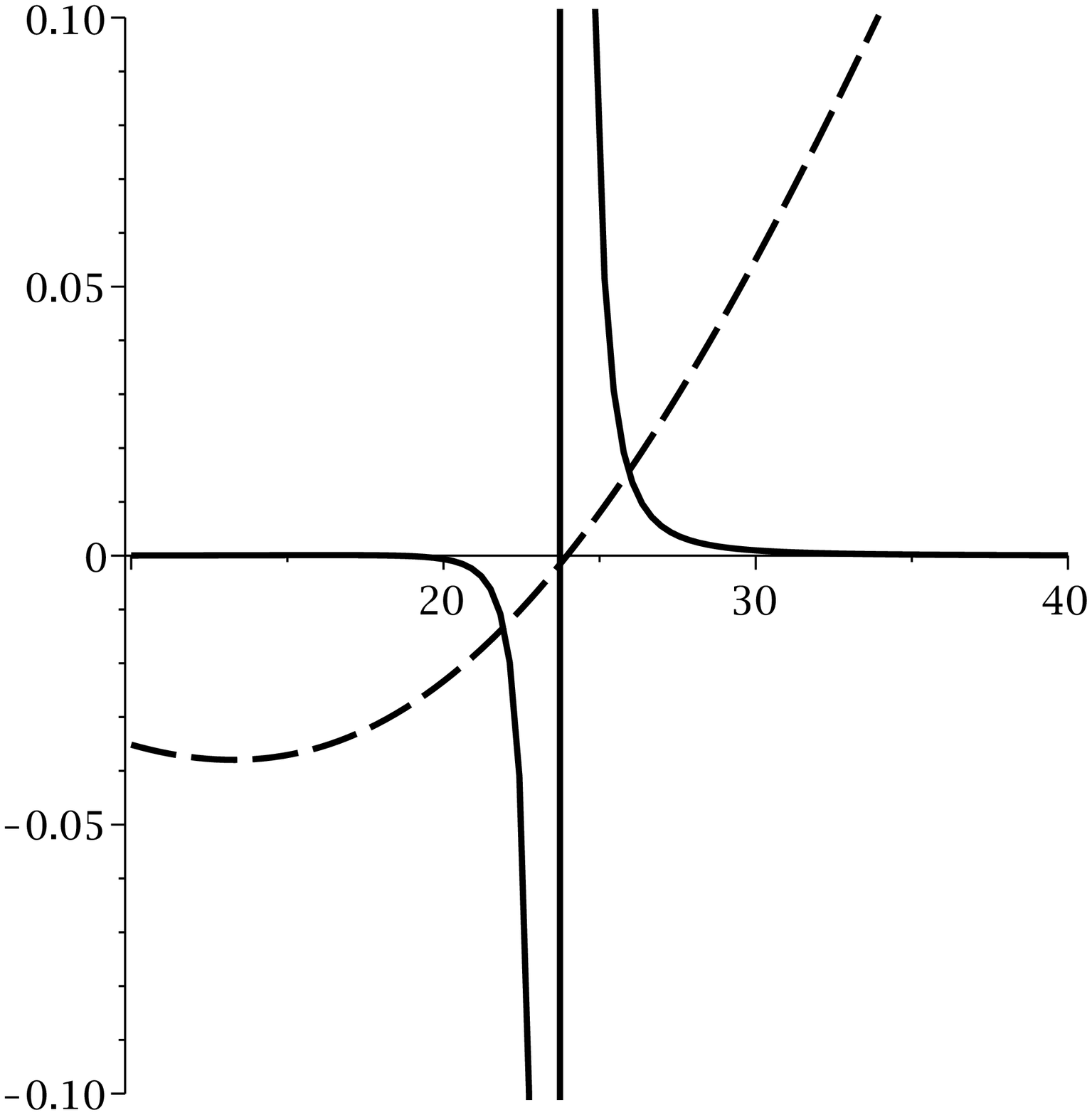}%
\end{array}
$%
\caption{$\mathcal{R}$ (continues line) and $C_{Q}$ (dashed line) versus $%
r_{+}$ for $l=1$, $\Lambda =-1$, $n=5$ and $q=1$, $\protect\alpha=2$, $%
b=0.05 $ and $k=1$. \newline
up diagrams for different scale: The Quevedo metric for case II. \newline
down diagrams for different scale: The Quevedo metric for case I. }
\label{Q}
\end{figure}

As for Quevedo metric case $II$, due to terms $M_{QQ}^{2}$ and $M_{\alpha
\alpha }^{2}$, there may be divergencies which may not be related to any
phase transition point of the heat capacity. In other words, these two terms
may have roots and their roots will contribute to number of divergencies of
the Ricci scalar (Fig. \ref{Q} up diagrams). In case of Quevedo case $I$, $%
SM_{S}+QM_{Q}+\alpha M_{\alpha }$ may have contribution to number of
divergencies of the Ricci scalar of this metric. The existence of these
divergencies are seen through following plotted graphs (Fig. \ref{Q} down
diagrams).

\begin{figure}[tbp]
$%
\begin{array}{cc}
\epsfxsize=6cm \epsffile{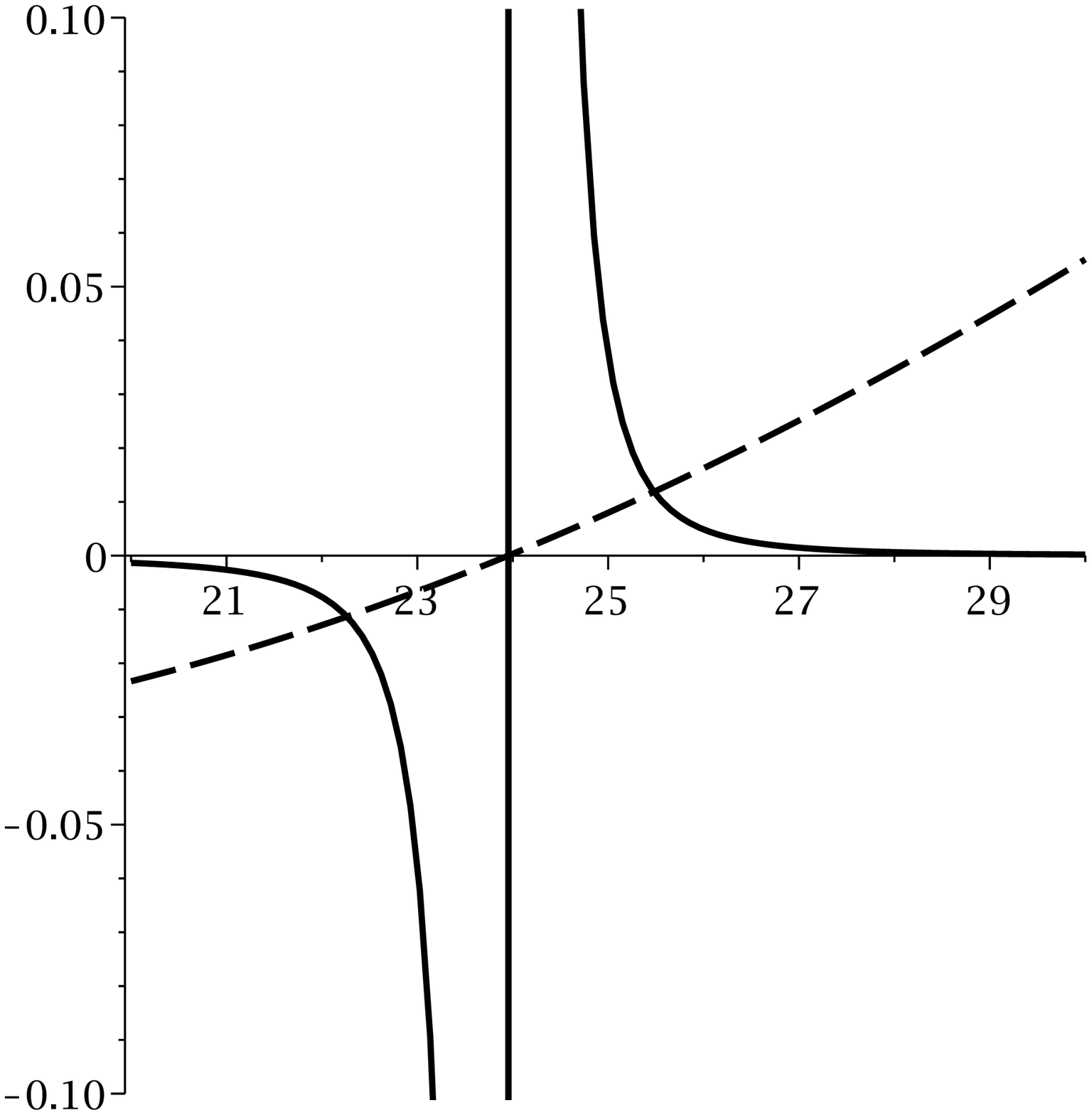} & \epsfxsize=6cm %
\epsffile{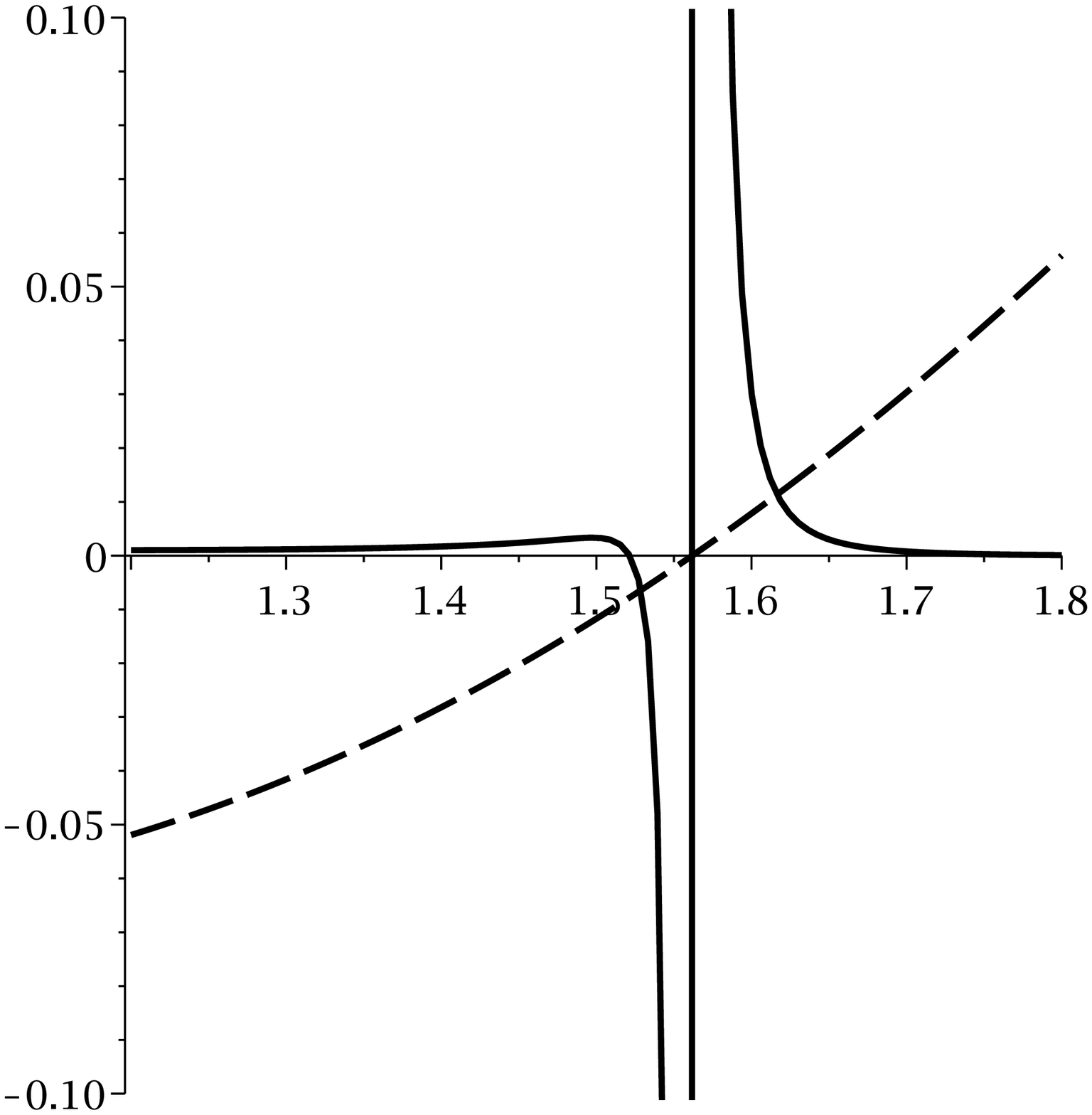}%
\end{array}
$%
\caption{$\mathcal{R}$ (continues line) and $C_{Q}$ (dashed line) versus $%
r_{+}$ for $l=1$, $\Lambda =-1$, $n=5$ and $q=1$, $\protect\alpha=2$ and $k=1
$. \newline
left diagram: $b=0.05$, right diagram: $b=0.5$. }
\label{New11}
\end{figure}

Finally, in case of HPEM metric, the denominator of the Ricci scalar of this
metric constructed in a way that there is no extra divergence point for its
Ricci scalar and all divergencies of the Ricci scalar and phase transition
points of the heat capacity will coincide. Therefore, this formalism is
providing a machinery with consisting results of studying heat capacity
(Fig. \ref{New11}).

\section{Closing Remarks}

In this paper, we have considered the thermodynamical behavior of the
topological charged dilaton black holes of EMd gravity. We have investigated
the phase transition points related to heat capacity and GTs of the EMd
black holes. We have shown that according to the variation of parameter
under consideration, the type and number of the phase transitions, stability
conditions, thermodynamical behavior of these black holes near critical
points will be modified. As for phase transitions, it was shown that these
black holes enjoys three thermodynamical behavior near critical points which
are nonphysical stable to physical stable, larger/smaller and smaller/larger
black holes. We have also pointed out that considering HPEM metric for
constructing thermodynamical spacetime leads to effective machinery for
studying phase transition points of the heat capacity. In other words,
divergencies of the Ricci scalar coincided with phase transition points of
heat capacity. It was also shown that Ricci scalar behavior near these
critical points depends on thermodynamical behavior near critical point.
Therefore, one is able to recognize the type of phase transition and its
behavior only by employing HPEM metric.

Next, we considered dilaton parameter as an extensive parameter and extended
thermodynamical space. We conducted an study with this consideration for
different approaches for GTs. We showed that in order to GTs methods and
heat capacity have consisting results, certain conditions must be satisfied
otherwise there might be extra divergencies. Only HPEM method was free of
any conditions. Later, It was shown that Weinhold, Ruppeiner and Quevedo
methods and their Ricci scalars contain divergencies which were not related
to any phase transition points of heat capacity. In other words, extra
divergencies were seen which were not consistent with results that were
obtained in case of heat capacity. Therefore, the mentioned conditions were
not hold for these pictures.

It is worthwhile to consider various models of nonlinear electrodynamics
instead of Maxwell field in this regard and examine the effects of
nonlinearity and also the validity of different thermodynamical metrics. We
left this issue for the forthcoming work.

\begin{acknowledgements}
We thank Shiraz University Research Council. This work has been
supported financially by the Research Institute for Astronomy and
Astrophysics of Maragha, Iran.
\end{acknowledgements}

\end{document}